\documentclass[twoside, letterpaper, 12pt]{article}

\DeclareMathSizes{10}{10}{10}{10}
\usepackage{anysize}
\usepackage{fancyhdr}
\marginsize{2cm}{2cm}{0.5cm}{0.75cm}
\usepackage{setspace}

\usepackage[utf8]{inputenc} 
\usepackage[T1]{fontenc}    
\usepackage{url}            
\usepackage{booktabs}       
\usepackage{amsfonts}       
\usepackage{nicefrac}       
\usepackage{microtype}      
\usepackage{amsthm}
\usepackage{subfigure}
\usepackage{caption}
\usepackage{subcaption}
\usepackage{graphics} 
\usepackage{epsfig} 
\usepackage{times} 
\usepackage{amsmath} 
\usepackage{amssymb}  
\usepackage{makeidx}
\usepackage{float}
\usepackage{balance}
\usepackage{booktabs}
\usepackage{bbm}
\usepackage{mathrsfs}
\usepackage{enumerate}
\usepackage{multicol}
\usepackage{algorithm}
\usepackage{algpseudocode}

\usepackage{mathrsfs}  
\usepackage[normalem]{ulem}
\usepackage{xr} 
\usepackage{multicol}
\usepackage{float}
\usepackage[nolist]{acronym}
\usepackage{xcolor}
\usepackage[colorlinks]{hyperref}       

\hypersetup{
  colorlinks = true,
  allcolors = siaminlinkcolor,
  urlcolor = siamexlinkcolor,
}
\colorlet{siaminlinkcolor}{green!50!black}
\colorlet{siamexlinkcolor}{red!50!black}

\usepackage{lipsum} 

\usepackage{setspace}

\usepackage{cleveref} 

\newcommand{\rd}{{\mathrm d}}

\newcommand{\vx}{{\bf x}}
\newcommand{\vy}{{\bf y}}
\newcommand{\vz}{{\bf z}}

\newcommand{\vw}{{\bf w}}

\newcommand{\vW}{{\bf W}}
\newcommand{\vX}{{\bf X}}
\newcommand{\vY}{{\bf Y}}

\newcommand{\calF}{{\cal F}}

\newcommand{\calH}{{\cal H}}

\newcommand{\calL}{{\cal L}}

\newcommand{\argmin}{\operatornamewithlimits{argmin}}

\newcommand{\T}{^\mathsf{T}}

\newcommand{\Cb}{\mathbb{C}}

\newcommand{\Nb}{\mathbb{N}}

\makeatletter
\renewcommand{\maketag@@@}[1]{\hbox{\m@th\normalsize\normalfont#1}}%
\makeatother

\catcode`\_=11\relax
    \newcommand\email[1]{\_email #1\q_nil}
    \def\_email#1@#2\q_nil{%
      \href{mailto:#1@#2}{{\emailfont #1\emailampersat #2}}
    }
    \newcommand\emailfont{}
    \newcommand\emailampersat{\small@}
    \catcode`\_=8\relax 
    
\makeatletter
\newcommand{\vast}{\bBigg@{3.5}}
\newcommand{\Vast}{\bBigg@{4}}
\newcommand{\vastt}{\bBigg@{4.5}}
\newcommand{\Vastt}{\bBigg@{5}}
\makeatother

\newcommand{\Uhat}{\hat{U}}
\newcommand{\Ihat}{\hat{I}}
\newcommand{\Ahat}{\hat{A}}
\newcommand{\Bhat}{\hat{B}}
\newcommand{\Ghat}{\hat{G}}
\newcommand{\Vhat}{\hat{V}}
\newcommand{\Xhat}{\hat{X}}
\newcommand{\Yhat}{\hat{Y}}
\newcommand{\yhat}{\hat{y}}
\newcommand{\Zhat}{\hat{Z}}
\newcommand{\Hhat}{\hat{H}}
\newcommand{\Rhat}{\hat{R}}
\newcommand{\ket}[1]{| #1 \rangle}
\newcommand{\bra}[1]{\langle #1 |}
\newcommand{\bracket}[2]{\langle #1 | #2 \rangle}
\newcommand{\kettext}[1]{\ket{\text{#1}}}

\newcommand\twovector[2]{%
\begin{bmatrix}
#1\\
#2 
\end{bmatrix}}
\newcommand\fourvector[4]{%
\begin{bmatrix}
#1\\
#2\\
#3\\
#4\\
\end{bmatrix}}
\newcommand{\bracketpsipsi}{\bracket{\psi}{\psi}}
\newcommand{\fraconeroottwo}{\frac{1}{\sqrt{2}}}
\newcommand{\subalpha}[1]{\alpha_{#1}}

\newcommand{\derivtheta}{\frac{\rd}{\rd \theta}}
\newcommand{\partialtheta}{\frac{\partial}{\partial \theta}}
\newcommand{\derivCtheta}{\frac{\rd C(\theta)}{\rd \theta}}
\newcommand{\brapsi}{\bra{\psi}}
\newcommand{\ketpsi}{\ket{\psi}}
\newcommand{\HthetaVcomm}{[\Hhat + \theta \Vhat, \cdot]}
\newcommand{\piovertwo}{\frac{\pi}{2}}

\newcommand{\pifour}{\frac{\pi}{4}}
\newcommand{\negpifour}{\frac{-\pi}{4}}
\newcommand{\halfsqrttwo}{\frac{\sqrt{2}}{2}}
\newcommand{\pztheta}{\frac{\partial Z(\theta)}{\partial \theta}}
\newcommand{\pptheta}{\frac{\partial}{\partial \theta}}
\newcommand{\half}{\frac{1}{2}}
\newcommand{\sixth}{\frac{1}{6}}
\newcommand{\twefourth}{\frac{1}{24}}
\newcommand{\ztheta}{Z(\theta)}

\newcommand{\beginsupplement}{
        \setcounter{table}{0}
        \renewcommand{\thetable}{A\arabic{table}}
        \setcounter{figure}{0}
        \renewcommand{\thefigure}{A\arabic{figure}}
        \setcounter{section}{0}
        \renewcommand{\thesection}{A\arabic{section}}
        \setcounter{equation}{0}
        \renewcommand{\theequation}{A\arabic{equation}}
        \setcounter{algorithm}{0}
        \renewcommand{\thealgorithm}{A\arabic{algorithm}}
     }

\begin{acronym}
\acro{DNN}{Deep Neural Network}
\acro{ODE}{Ordinary Differential Equation}
\acro{SPDE}{Stochastic Partial Differential Equation}
\acro{FNN}{Feed-forward Neural Network}
\acro{CNN}{Convolutional Neural Network}
\acro{DP}{Dynamic Programming}
\acro{LSTM}{Long-Short Term Memory}
\acro{FC}{Fully Connected}
\acro{DDP}{Differential Dynamic Programming}
\acro{HJB}{Hamilton-Jacobi-Bellman}
\acro{PDE}{Partial Differential Equation}
\acro{PI}{Path Integral}
\acro{NN}{Neural Network}
\acro{SOC}{Stochastic Optimal Control}
\acro{RL}{Reinforcement Learning}
\acro{MPC}{Model Predictive Control}
\acro{IL}{Imitation Learning}
\acro{RNN}{Recurrent Neural Network}
\acro{DL}{Deep Learning}
\acro{SGD}{Stochastic Gradient Descent}
\acro{SDE}{Stochastic Differential Equation}
\acro{VRL}{Variational Reinforcement Learning}
\acro{IDVRL}{Infinite Dimensional Variational Reinforcement Learning}
\acro{1D}{1-dimensional}
\acro{2D}{2-dimensional}
\acro{ROM}{Reduced Order Model}
\acro{SDE}{Stochastic \ac{ODE}}
\end{acronym}

\begin{document}
	
	\title{
	\rule{\linewidth}{4pt}\vspace{0.3cm} \Large \textbf{
        A Quick Introduction to Quantum Machine Learning \\ for Non-Practitioners
	}\\ \rule{\linewidth}{1.5pt}}
	\author{Dominic Byrne\thanks{Authors contributed equally. Names appear in alphabetical order.},\, Matthew G. Cook\footnotemark[1],\, and Ethan N. Evans\footnotemark[1]\;\,\thanks{Corresponding Author. Email: \email{ethan.n.evans.civ@us.navy.mil}}\,\, \\ \vspace{-0.1cm}
	\small{Naval Surface Warfare Center Panama City Division} \\ \vspace{-0.2cm}
	 }
	\date{}
	
	\maketitle

\begin{abstract}
Quantum computing has quickly become a highly active research area, and quantum machine learning has emerged as a potential manifestation of classical machine learning on quantum hardware. The widespread successes of classical machine learning in classification problems are extremely attractive, however they come at the cost of an exponential growth of parameters in modern network architectures (e.g. GPT). A possible benefit in addressing such problems with quantum networks is an increased expressibility of quantum bits over classical bits, which through quantum machine learning leads to an increased expressibility of a quantum neuron. 

Quantum computing is founded on the premise of using particles that are governed by quantum mechanics for the purposes of computation by leveraging key aspects such as superposition and entanglement. These properties have theoretical advantage in representing and manipulating information. Namely superposition allows for a fundamental bit of information to encode a continuous spectrum, while entanglement allows non-local effects to manipulate encoded information. Circuits of quantum gates are used to perform quantum computations, and when parameterized, can be optimized, or trained, using traditional methods in optimization. This leads to a quantum machine learning framework where classical information embedded in quantum bits can take advantage of quantum phenomena and increased expressibility for a potential reduction in network size and training time on quantum hardware. 

This manuscript serves as introductory material for researchers that are new to the areas of quantum mechanics and machine learning, in order to decrease the timeframe needed for developing new expertise. The notion of a Turing machine is used as a foundation and motivation for creating computers out of quantum hardware. Next, basic principles and notation of quantum mechanics are introduced, including superposition, phase space on the Bloch sphere, and entanglement of multiple quantum bits. A basic review of classical digital logic is used to propose notions of quantum gates that may leverage these key properties by a universal set of quantum gates. Next, we introduce classical deep learning concepts such as the artificial neural network, the gradient descent algorithm and its stochastic generalization, and the standard backpropagation approach to training a neural network. These are used as a foundation for introducing trainable quantum circuits as neural networks, including a derivation of the analogous gradient descent approach and its generalizations and methods of encoding classical information in a quantum circuit. Finally, these are topics are combined in an illustrative example problem that highlights a potential advantage of quantum neural networks. The accompanying appendices offer greater detail of various derivations that are provided throughout the manuscript. 

\end{abstract}


\section{Why Quantum Computing}\label{sec:intro}
The notion of a universal computer was first characterized by the description of the Turing machine \cite{turing2004intelligent} based on an infinitely long tape, or computation register, which through the machine could be used to encode any algorithm. These notions, along with the advent of the transistor, eventually led to the von Neumann architecture of computing we have today, where a central processing unit (CPU) performs sequential operations on an encoded stream of information, and separately stores this information in a memory register. With the trend of increasing transistor density per unit area, eventually a limit could be perceived. In addition, the polynomial time computational complexity constraint of this computing architecture, as suggested by the Church-Turing thesis, made the quest for alternative architectures inevitable \cite{naja2022development}. 

The idea of leveraging quantum physics to perform computations was first suggested by Feynman, and is by no means the only alternative to the von Neumann architecture (see \cite{adamatzky2016advances,adamatzky2005reaction,shastri2021photonics}), however its appeal is based on the ability to leverage attractive properties such as superposition and entanglement, which allow for a unique framework. Herein, one must carefully construct algorithms which yield a choreography of operations that use sequential constructive and destructive interference to promote the correct solution in probability while effectively cancelling out the probability of incorrect solutions, all without a-priori knowledge of the correct solution. In principle, this approach promises an exponential speedup over the von Neumann architecture \cite{aaronson2022introduction}. 

The novelty and unique challenges of leveraging quantum physics for computing has led to not just one, but numerous computing architectures explored by a growing contingent of companies and organizations, each seeking to establish their approach as the dominant approach, and to be the first to demonstrate quantum supremacy on a large-scale device. Among these are: a) using superconducting qubits (often referred to as SQUIDs) by IBM, Google, USTC, and Rigetti, b) trapped ions by IonQ and others, c) photonics by Xanadu and others, and trapped Rydberg atoms by QuEra and others. Beyond these hardware implementations, are several other floating notions, such as building quantum computing on qudits (d-level quantum bits, e.g. ternary quantum bit) as opposed to the binary quantum bit (qubit) \cite{wang2020qudits}, as well as abandoning the discrete framework altogether and instead building computing on a continuum quantum state \cite{lloyd1999quantum,pfister2019continuous}. 

Outside the goal of universal quantum computing are also specific quantum realizations for specific applications, as in the case of quantum annealing \cite{kadowaki1998quantum,hauke2020perspectives} for optimization problems, and quantum machine learning \cite{naja2022development,biamonte2017quantum} for learning problems. The rapid and heavy investment in quantum computing has led to a so-called race for quantum supremacy in which proponents suggest it can be achieved as soon as the early 2030s \cite{google2023our}, yet there remain substantial hurdles before this dream may be realized. 


\section{Quantum Computing Overview}\label{sec:overview}


\subsection{Quantum Physics - Notation and Introduction}
Quantum computing relies on the manipulation and measurement of quantum phenomena in order to process information. The behavior of quantum phenomena is described by quantum physics, so an understanding of quantum computing necessitates a basic understanding of quantum physics. Quantum physics introduces math notation, known as Dirac notation, which is often quite foreign to other disciplines yet simplifies the introduction of main concepts and carries over to quantum computing. Dirac notation utilizes linear algebra, probability and statistics, as well as notational conventions used in physics. 

Linear algebra is used in Dirac notation to describe quantum states as vectors, and to describe the physical processes that can impart change to a quantum state as matrices\footnote{In the more general setting, an operator can act on an infinite vector space, but this brief introduction limits vectors to finite dimensional vector spaces where operators are simply described by standard matrices}, which in quantum literature are referred to as operators. Operators ‘operate’ on these quantum states to give some new quantum state. A column vector is by convention used to represent some physical quantum state, for example whether an electron is in the spin up state or spin down state or a photon is in the horizontal or vertical linear polarization state. A vector state $\psi$ (the variable $\psi$ is often used in literature to describe a generic quantum state, usually called a ‘wavefunction’) with two quantities would normally be written as: 
\begin{equation}
    \vec{\psi} = \begin{bmatrix}
         a\\
         b 
    \end{bmatrix}.
\end{equation}
In Dirac notation, the vector state $\psi$ would be written as:
\begin{equation}
        |\psi\rangle = \begin{bmatrix}
         a\\
         b 
    \end{bmatrix},
\end{equation}
where $|\cdot\rangle$ is the right side of a bracket $\langle \cdot | \cdot \rangle$ and is called a ``ket'' vector. The complex conjugate transpose of a vector state is also widely used, and is written as: 
\begin{equation}
    \vec{\psi}^{*\mathsf{T}} = [a^* \;\;\; b^*] = \vec{\psi}^\dagger,
\end{equation}
where $*$ is the complex conjugate, $\cdot\T$ is the transpose, and the dagger $\cdot^\dagger$ is shorthand notation in quantum physics for the conjugate transpose. In Dirac notation, the complex conjugate transposed of a vector state $\psi$ is written as: 
\begin{equation}
    \langle \psi | = [a^*\;\;\; b^*],
\end{equation}
where $\langle \cdot |$ is the left side of the bracket $\langle \cdot | \cdot \rangle$ and is called a ``bra'' vector. A bra and a ket written next to each other in a full bracket denotes an inner product between the two vector states:
\begin{equation}
    \langle \psi_1 | \psi_2 \rangle = \vec{\psi_1}^\dagger \cdot \vec{\psi_2} = [a_1^* \;\;\; b_1^*] \cdot \begin{bmatrix}
         a_2\\
         b_2 
    \end{bmatrix} = a_1^*a_2 + b_1^* b_2.
\end{equation}
Unitary operators, represented by matrices, are used to change the quantum state. For example,
some state $\psi_i$ can be modified to another state, $\psi_{i+1}$ after the application of an operator $\hat{U}$ (the hat $\hat{\cdot}$ symbol is often used to denote an operator, though it is not always used). Operators are
usually given capital letters for variable names, and a unitary operator is a special type of operator that satisfies the property:
\begin{equation}
    \Uhat \Uhat^\dagger = \Uhat^\dagger \Uhat = \Ihat,
\end{equation}
where $\Ihat$ is the identity operator and corresponds to imparting \textit{no change} to a quantum state. The
action of an operator, represented as a matrix, is written as:
\begin{equation}
    \ket{\psi_{i+1}} = \Uhat \ket{\psi_i} = \begin{bmatrix}
        u_{11} & u_{12} \\
        u_{21} & u_{22}
    \end{bmatrix} \begin{bmatrix}
        a_i \\
        b_i
    \end{bmatrix} = \begin{bmatrix}
        u_{11} a_i + u_{12} b_i \\
        u_{21} a_i + u_{22} b_i
    \end{bmatrix} = \begin{bmatrix}
        a_{i+1} \\
        b_{i+1}
    \end{bmatrix}.
\end{equation}

One can similarly apply complex conjugates and transposes to an operator,
\begin{equation}
    \Uhat^\dagger = \begin{bmatrix}
        u_{11}^* & u_{21}^* \\
        u_{12}^* & u_{22}^*
    \end{bmatrix},
\end{equation}
and the dagger of an operator is typically applied to bra vectors to update them:
\begin{equation}
    \bra{\psi_{i+1}} = \bra{\psi_i}\Uhat^\dagger = [a_i^* \;\;\; b_i^*]\begin{bmatrix}
        u_{11}^* & u_{21}^* \\
        u_{12}^* & u_{22}^*
    \end{bmatrix} = [a_i^*u_{11}^* + b_i^*u_{12}^*\;\;\; a_i^*u_{21}^* + b_i^* u_{22}^*] = [a_{i+1}^*\;\;\; b_{i+1}^*].
\end{equation}

Some operators are used to represent measurable properties of quantum systems and are called
‘observables’ or ‘observable operators’. These operators satisfy a slightly weaker requirement of
being Hermitian. Hermitian operators are defined by satisfying the following relation:
\begin{equation}
    \Ahat = \Ahat^\dagger .
\end{equation}
An important property of Hermitian operators is that one can easily show that they have all real eigenvalues despite having potentially all complex entries. Furthermore, Hermitian operators typically represent ‘measurable’ operators, and in such cases their eigenvalues correspond to
familiar quantities such as position, momentum, energy, etc., which also must be real-valued.

The average value of all possible outcomes, based on the probability of each outcome, is referred to as the expectation value (or expected value) of that observable with respect to that quantum state. In Dirac notation, the expectation value is expressed as:
\begin{equation}
    \langle \psi | \Ahat | \psi \rangle,
\end{equation}
where $\Ahat$ is some observable (which we represent here by a matrix), and $\ket{\psi}$ is the quantum state
just before the state is measured. The expectation value describes the average measurement outcome, so despite the discrete values of single measurements, an expectation value can (and usually will) give a non-discrete value that does not correspond to a single possible measurement outcome. The possible discrete-valued outcomes of each measurement are given by the eigenvalues of observable $\Ahat$. 

Performing the calculation above will yield the average of the possible outcomes weighted by the probability of each outcome occurring. Expectation values can also be estimated empirically by performing repeated measurements and averaging the results. The probability will be inherent in the empirical histogram of each of the possible outcomes, and if measured an infinite number of times would result in the exact expectation value. Since infinite measurements are not practically feasible, an estimate for the expectation value can be obtained by a finite number of measurements and is often treated as the empirical expectation.

Another important calculation in quantum physics is the probability of some general quantum state $\ket{\psi}$ being in the specified, known state $\ket{\phi}$. This is calculated by the magnitude squared of
the inner product between $\ket{\psi}$ and $\ket{\phi}$, which in Dirac notation is represented as:
\begin{equation}
    |\bracket{\phi}{\psi}|^2 = (\bracket{\phi}{\psi})^\dagger\bracket{\phi}{\psi} = \bracket{\psi}{\phi}\bracket{\phi}{\psi}
\end{equation}
The inner product $\bracket{\phi}{\psi}$ is the projection of the state $\ket{\psi}$ onto the desired measurement state $\ket{\phi}$. The inner product describes the ‘amount of overlap’ due to the projection, and is called the
amplitude of the probability or the ‘probability amplitude’. The square of the magnitude of the probability amplitude (which can also be computed by multiplication of the inner product with its complex conjugate), gives a real value that corresponds to the probability of finding the state $\ket{\psi}$ in the state $\ket{\phi}$ when measured. 

In quantum computing, most calculations are performed in the computational basis formed by the zero state $\ket{0}$ and the one state $\ket{1}$ where:
\begin{equation}
    \begin{array}{l c r}
        \ket{0} = \twovector{1}{0} & \text{and} & \ket{1} = \twovector{0}{1}.
    \end{array}
\end{equation}
These states are unit vectors because they have a norm (or magnitude, computed as the inner
product of a state with itself) of one: 
\begin{equation}
    \begin{array}{l c r}
        \bracket{0}{0} = 1 & \text{and} & \bracket{1}{1} = 1.
    \end{array}
\end{equation}
They are also orthogonal to each other, meaning inner products with each other yield zero:
\begin{equation}
    \begin{array}{l c r}
        \bracket{0}{1} = 0 & \text{and} & \bracket{1}{0} = 0.
    \end{array}
\end{equation}
Finally, the basis is ‘complete’, meaning that any arbitrary vector can be written as a linear
combination of these two basis elements:
\begin{equation}
    \ket{\psi} = \twovector{a}{b} = a \twovector{1}{0} + b \twovector{0}{1}
\end{equation}
Together, these three properties describe the basis as being a complete orthonormal basis.

So far these states seem similar to the binary zeros and ones that are used to represent values in
classical computing, however quantum mechanics offers more possible states, which quantum
computing seeks to utilize. While
\begin{equation}
    \begin{array}{l c r}
        \ket{\psi} = \twovector{1}{0} & \text{and} & \ket{\psi} = \twovector{0}{1}
    \end{array} 
\end{equation}
would be the only possible states in a classical computing scheme, the state
\begin{equation}
    \ket{\psi} = \twovector{1}{1}
\end{equation}
is a legitimate, though unnormalized, state in quantum mechanics. Normalization factors need to
be added to ensure state probabilities sum to unity, which will be elucidated by an example. First,
the importance of this state being allowed should be highlighted. The state $\ket{\psi} = \twovector{1}{1}$ is a linear combination of the zero state and the one state:
\begin{equation}
    \ket{\psi} = \twovector{1}{1} = \twovector{1}{0} + \twovector{0}{1} = \ket{0} + \ket{1},
\end{equation}
and represents some quantum state at a certain point or time before measurement. The fact that
multiple possible states are present simultaneously is called superposition, and is a property
unique to quantum mechanics. This superposition of states is itself a quantum state; operations
and gates applied to the state $\ket{\psi}$ are applied in their normal manner and act on both parts of the
state present in the superposition state simultaneously. This key quantum phenomenon leads to a
larger computational space that quantum computing seeks to leverage; the classical binary
representation is replaced with a continuous space of possible superposition states.

However, once a measurement is made only one of the possible measurable values remains,
which was predicted with some probability. So before measurement, both quantum states exist
simultaneously in the quantum superposition state, but upon measurement this quantum state
``collapses'', and only one basis state is observed while the information for the other state is lost.
This probability is mathematically determined by the complex coefficients that multiply each
basis element, which are also used to normalize the quantum state. 

Now to show why those coefficients are needed, a counter example is presented. Starting with
the unnormalized superposition state (without coefficients) introduced earlier:
\begin{equation}
    \ket{\psi} = \ket{0} + \ket{1} = \twovector{1}{0} + \twovector{0}{1} = \twovector{1}{1}.
\end{equation}
If we take its norm
\begin{equation}
    \bracket{\psi}{\psi} = [1^* \;\;\; 1^*] \twovector{1}{1} = 2,
\end{equation}
we obtain a value of 2. The issue with this becomes apparent when the magnitude of the inner
product is squared:
\begin{equation}
    |\bracketpsipsi|^2 = (\bracketpsipsi)^\dagger \bracketpsipsi = \bracketpsipsi\bracketpsipsi = 2 \cdot 2 = 4
\end{equation}

As described earlier, the magnitude of the inner product of two states squared gives the
probability that one state will be in the other state when measured. If the same operation is
performed on a state with itself, it yields the probability of the state being in its own state, which
must yield a probability of 1 or 100\%. Without any normalization coefficients, this calculation
gives a probability of 4, or 400\%, which is a nonsensical and nonphysical result. Therefore, a
condition is imposed that the magnitude of the norm of a state squared must always give a
probability of one to ensure the physics remains consistent and logical: 
\begin{equation}
    |\bracketpsipsi|^2 \equiv 1
\end{equation}
This condition also implies that
\begin{equation*}
    |\bracketpsipsi|^2 = (\bracketpsipsi)^\dagger \bracketpsipsi = \bracketpsipsi\bracketpsipsi = (|\bracketpsipsi|^2 = (\bracketpsipsi)^\dagger \bracketpsipsi = \bracketpsipsi\bracketpsipsi)^2 \equiv 1,
\end{equation*}
and
\begin{equation}
    \bracketpsipsi \equiv 1.
\end{equation}

Coefficients are added to a state to ensure this condition remains true. This is often referred to as
normalization. For the above superposition state: 
\begin{equation}
    \ket{\psi} = \twovector{a}{b} = a\ket{0} + b \ket{1} = a\twovector{1}{0} + b\twovector{0}{1},
\end{equation}
this condition gives constraints on $a$ and $b$:
\begin{equation}
        \bracketpsipsi = [a^* \;\;\; b^*]\twovector{a}{b} = a^*a + b*b = 1.
\end{equation}
The coefficients $a$ and $b$ are also used to describe the probability of measuring the state
associated with that coefficient. For example, say a state $\ket{\psi}$ has coefficients $a=1/\sqrt{3}$ and $b=\sqrt{2/3}$:
\begin{equation}
    \ket{\psi} = \frac{1}{\sqrt{3}}\ket{0}+\sqrt{\frac{2}{3}}\ket{1} = \frac{1}{\sqrt{3}}\twovector{1}{0}+\sqrt{\frac{2}{3}}\twovector{0}{1} = \twovector{\frac{1}{\sqrt{3}}}{\sqrt{\frac{2}{3}}}.
\end{equation}

To calculate the probability of measuring the zero state from this superposition state, take
the square of the magnitude of the inner product of this state with the zero state:
\begin{equation}
    \begin{split}
        |\bracket{0}{\psi}|^2 &= \left[\frac{1}{\sqrt{3}}\;\;\; \sqrt{\frac{2}{3}}\right]\twovector{1}{0}[1\;\;\; 0]\twovector{\frac{1}{\sqrt{3}}}{\sqrt{\frac{2}{3}}} \\
        &= \left[\frac{1}{\sqrt{3}}\;\;\; \sqrt{\frac{2}{3}}\right] \begin{bmatrix}
            1 & 0 \\
            0 & 0
        \end{bmatrix} \twovector{\frac{1}{\sqrt{3}}}{\sqrt{\frac{2}{3}}} = \left[\frac{1}{\sqrt{3}}\;\;\; \sqrt{\frac{2}{3}}\right]\twovector{\frac{1}{\sqrt{3}}}{0} = \frac{1}{3}
    \end{split}
\end{equation}
This means there is a $1/3$ probability of measuring the eigenvalue associated with the “zero”
state. Performing a similar calculation for the one state yields $2/3$, meaning there is a $2/3$ probability of measuring the eigenvalue associated with the one state. 

It is also important to note that the magnitude squared of $a$ and $b$ gives the respective
probabilities of each state’s eigenvalues being measured:
\begin{equation}
    \begin{array}{l c r}
        a^*a = \frac{1}{3} & \text{and} & b^*b = \frac{2}{3}.
    \end{array}
\end{equation}
This can be seen clearly by doing the probability calculations for measuring the zero state and
the one state with a generic state $\ket{\psi} = a \ket{0} + b\ket{1}$ where $a$ and $b$ are left as variables. 
\newline Zero state:
\begin{equation}
    \begin{split}
        |\bracket{0}{\psi}|^2 &= \left[a^*\;\;\; b^*\right]\twovector{1}{0}[1\;\;\; 0]\twovector{a}{b} \\
        &= \left[a^*\;\;\; b^*\right] \begin{bmatrix}
            1 & 0 \\
            0 & 0
        \end{bmatrix} \twovector{a}{b} = \left[a^*\;\;\; b^*\right]\twovector{a}{0} = a^*a
    \end{split}
\end{equation}
\newline One state:
\begin{equation}
    \begin{split}
        |\bracket{1}{\psi}|^2 &= \left[a^*\;\;\; b^*\right]\twovector{0}{1}[0\;\;\; 1]\twovector{a}{b} \\
        &= \left[a^*\;\;\; b^*\right] \begin{bmatrix}
            0 & 0 \\
            0 & 1
        \end{bmatrix} \twovector{a}{b} = \left[a^*\;\;\; b^*\right]\twovector{0}{b} = b^*b
    \end{split}
\end{equation}
For further reading, see references \cite{townsend2010quantum,zettili2009quantum,le2011quantum,nielsen2010quantum,kaye2006introduction}.

\subsection{Superposition and Entanglement}

The two fundamental properties of quantum physics that are leveraged by quantum computing
are superposition and entanglement. When operating with qubits, oftentimes the pertinent
physical property being leveraged for quantum computing is known as the spin state. Let $\ket{s}$
represent the spin state of a 2-state particle (i.e. spin-up and spin-down). Then superposition is
simply an expression of the spin state in terms of its basis elements
\begin{equation}
    \ket{s} = \alpha \kettext{up} + \beta \kettext{down},
\end{equation}
where $\alpha, \beta \in \Cb$ are complex numbers such that the state is normalized, i.e. $|\alpha|^2 + |\beta|^2 = 1$, $\ket{\text{up}}$ represents the particle being spin-up, and $\kettext{down}$ represents the particle being spin-down. At a given time, the particle is in a linear combination of its two basis states, which is also
referred to as being in a superposition of its basis states, as described earlier. Upon measurement,
the superposition collapses to a single state outcome, which is often referred to as ‘wave function
collapse’.

\begin{figure}[H]
    \centering
    \includegraphics[width=0.5\textwidth]{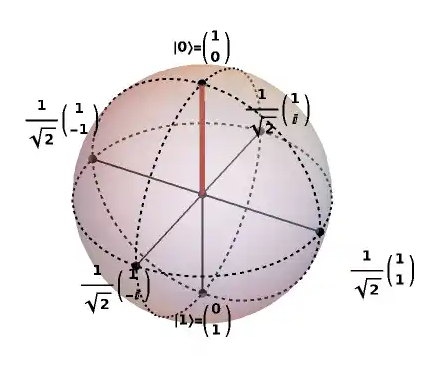}
    \caption{Bloch sphere representation of a qubit. This figure originated from \href{https://demonstrations.wolfram.com/QubitsOnThePoincareBlochSphere/}{https://demonstrations.wolfram.com/QubitsOnThePoincareBlochSphere/}}
    \label{fig:bloch_sphere}
\end{figure}

While this notion is quite simple from the perspective of linear algebra, it is far more interesting
from the perspective of probability theory. As explored earlier, these complex coefficients of
superposition are closely related to the respective probabilities of each outcome. For this reason,
often times they are referred to as probability amplitudes. The association to probabilities in the
classical sense comes by squaring these amplitudes, namely
\begin{equation}
    \begin{array}{cc}
        P(\kettext{up}) = |\alpha|^2, & P(\kettext{down}) = |\beta|^2. 
    \end{array}
\end{equation}
Thus, if for the moment, we take these amplitudes to be purely real, assign the $\kettext{up}$ state to be a
2-vector of unit length pointing in a positive-z direction, and assign the $\kettext{down}$ state to be a 2-
vector of unit length pointing in a negative-z direction, then superposition describes a 2D circle of unit length since we require $|\alpha|^2 + |\beta|^2 = 1$. The vector orthogonal to the z-axis in this circle corresponds to a state $\ket{s} = \fraconeroottwo\kettext{up} + \fraconeroottwo\kettext{down}$, where each outcome has equal probability, and this direction is assigned positive-x. Now, returning to the more general case of complex probability amplitudes, the imaginary direction creates a new y axis, and our 2D circular representation becomes a 3D sphere of unit length. This representation is called the Bloch sphere representation, and is depicted in \cref{fig:bloch_sphere}.

The property of entanglement is more complex, but also emerges mathematically in a deceptively simple form. Essentially, if one has two 2-state spin particles
\begin{equation}
    \begin{array}{l r}
         \ket{s_1} = \fraconeroottwo\kettext{up} + \fraconeroottwo\kettext{down} & \ket{s_2} = \fraconeroottwo\kettext{down} + \fraconeroottwo\kettext{up}
    \end{array}
\end{equation}
and can express their combined state $\ket{s_1,\, s_2}$ as
\begin{equation}
    \ket{s_1,\, s_2} = \fraconeroottwo \kettext{up,\, down} + \fraconeroottwo \kettext{down,\, up},
\end{equation}
then the two particles are said to be in a state of maximal entanglement. That is, since the total two-particle state is merely expressed by these two two-particle basis states, their states are coupled in such a way that if one were to measure the state of one of the particles (say in the up state), then they would have complete information of the state of the other (which must be in the down state). We say the particles are maximally entangled when each outcome has equal probability (50\% for this example). For a two qubit system, there are four such maximally entangled states, known as Bell states. Since the act of measurement of one particle gives complete information of the other, entanglement is a non-local phenomenon, so that virtually all
manipulations (i.e. through computing) of the state of one of the particles has some effect on the other particle, regardless of how separated the two particles are.

It is however not the case that all multi-particle systems are entangled. If one can separate the total system state into products of subsystem states, then the state is not entangled. This can be depicted in the following way. Say we have the two-particle system state:
\begin{equation}
    \ket{s_1,\, s_2} = \alpha \kettext{up,\, down} + \beta \kettext{up,\, up},
\end{equation}
This system is not in an entangled state because it can be expressed as a product of subsystem states as:
\begin{equation}
    \ket{s_1,\, s_2} = \kettext{up}\big(\alpha \kettext{down} + \beta \kettext{up}\big),
\end{equation}
Thus, if we were to measure the first particle, which must be in an up state, we would not gain any information about the state of the second particle, which could still be either in the up or the down state.


\subsection{Fundamentals of Classical Computing}

\begin{figure}[H]
    \centering
    \includegraphics[width=0.6\textwidth]{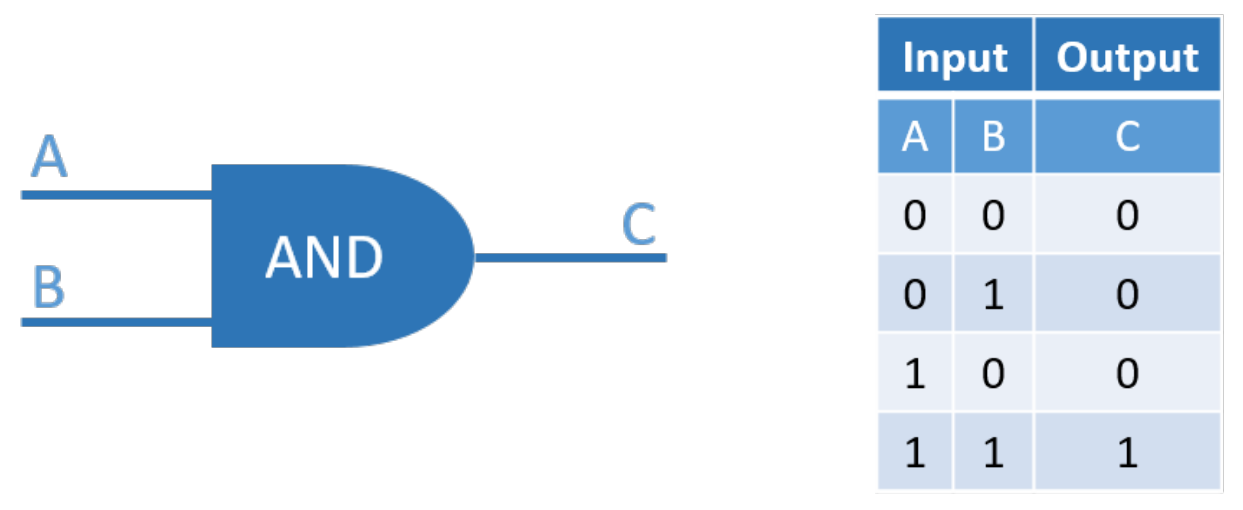}
    \caption{The AND gate}
    \label{fig:and_gate}
\end{figure}

In classical computing, we construct every computation from a fundamental set of building blocks: logical gates. These gates act on pairs of bits by conditionally flipping, summing, or otherwise manipulating the bit pair to yield a single output bit. A simple example is the AND gate shown in \cref{fig:and_gate} The AND gate returns 1 if both A and B are 1, and returns 0 otherwise, which is analogous to multiplying the inputs. The opposite of the AND gate is the NAND (NOT AND) gate, which returns 1 if both A and B are 0, and 0 otherwise. In contrast, the OR gate returns 1 if either A or B is 1, and 0 otherwise, which is analogous to addition. The NOR (NOT OR) gate is its opposite, returning 1 if either A or B is 0, and 1 otherwise. Another important gate
is the XOR gate, which returns 1 if A and B are different, and 0 if they are the same, analogous to an equality test. Its opposite, the XNOR gate, returns 0 if A and B are different, and returns 1 if they are the same.

\begin{figure}[H]
    \centering
    \includegraphics[width=0.7\textwidth]{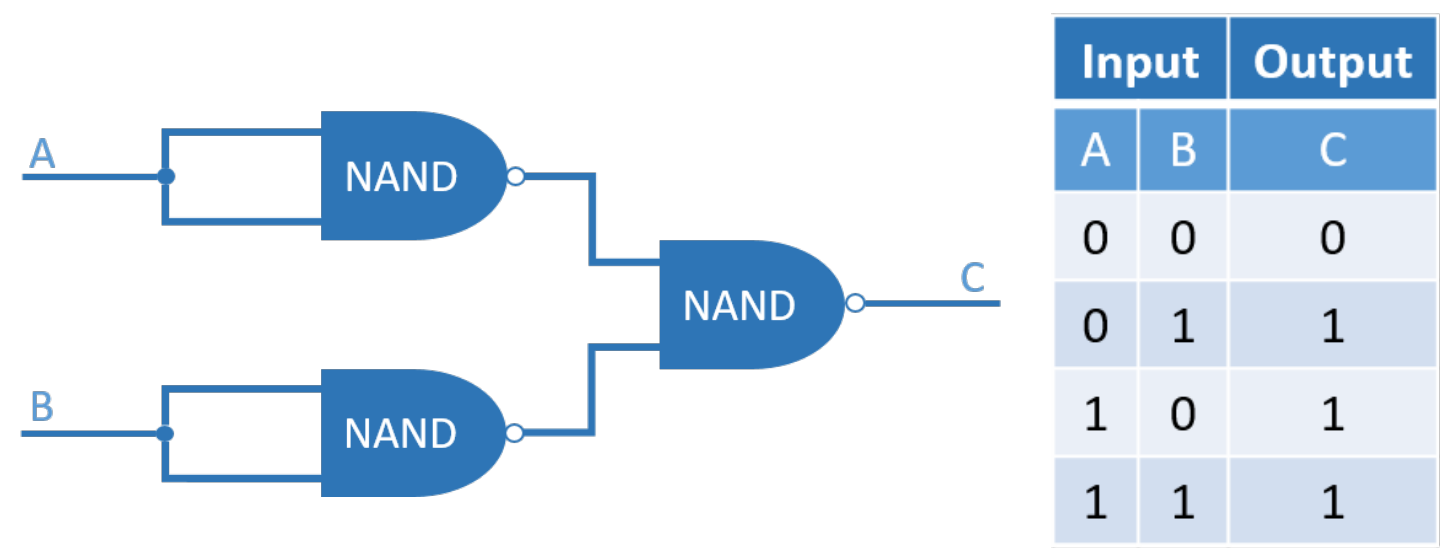}
    \caption{The OR gate as a circuit of only NAND gates.}
    \label{fig:or_gate}
\end{figure}

These 6 gates may be applied in a sequence, referred to as a circuit, to create complex networks of logical operations that in turn may be used to construct any algorithm or set of operations. Connecting back to the ideas of Turing, these logical gate building blocks form a universal computer, and are the foundation of modern digital computing. An interesting thing to note is that many of these gates can be created by circuits of a different gate. For example, as depicted in \cref{fig:or_gate}, one can construct an OR gate from a circuit of only NAND gates. Actually, one can construct any gate with a circuit of only NAND gates. Similarly, one can construct any gate with a circuit of only NOR gates. We refer to each of these individually as their own universal gate basis set. This is an important notion that will be explored shortly.

\subsection{Fundamentals of Quantum Computing}

Similar to the notion of a classical gate, the main thread of quantum computing research seeks to compose quantum computers out of similarly defined logic gates and circuits of logic gates in order to rebuild the classical computing architecture we are familiar with today from the groundup, and thus produce a universal computer out of quantum components. Instead of acting on bits, these components act on quantum bits, or qubits, which are binary representations of quantum states in a 2-state system.

Consider again the spin state of an electron. This state can either be spin-up or spin-down. We previously used the ket notation to describe this as the $\kettext{up}$ state and the $\kettext{down}$ state, but equivalently, one can write them, respectively, as $\ket{0}$ and $\ket{1}$, so that we have encoded the two possible spin state outcomes into a binary representation. These encoded binary representations both simplify our notation, but also generalize to cases where the primary leveraged property is not spin, but some other 2-state property (e.g. the excited/ground state of a Rydberg atom). 

The fundamental difference in how this binary state behaves is again the notion of superposition.
When each qubit is observed (measured), it must be either in the $\ket{0}$ or the $\ket{1}$ state, however
during the sequence of gate operations, it is described by a superposition state
\begin{equation}
    \ket{s} = \alpha\ket{0} + \beta\ket{1}.
\end{equation}
Thus, similarly, our quantum logic gates must be able to act not only on binary states, but also on
any superposition of binary states. Since we always describe states with respect to this basis, it is
convenient to write the state vector only in terms of the probability amplitudes $\alpha$ and $\beta$ as
\begin{equation}
    \ket{s} = \twovector{\alpha}{\beta},
\end{equation}
where by convention the top entry corresponds to the amplitude of the $\ket{0}$ basis state and the
bottom entry corresponds to the amplitude of the $\ket{1}$ basis state.

Another complication arises from the time-reversibility of quantum mechanics. This requirement
states that one must be able to run any sequence of operations either forwards or backwards.
Mathematically, this means that every gate operation must be \textit{one-to-one}, that is, an operation on
some unique input produces some unique output. Thus each quantum gate operation must be
\textit{invertible}. This is not true of classical gates; a variety of different inputs can produce the same
output, and furthermore, the size of the input space and the size of the output space need not
match. This can be clearly seen in the AND logic gate diagram. Two inputs produce a single
output, and the output is not unique to the set of inputs (e.g. there are several input pairs that
produce 0). Thus, since we have a binary representation, and our input and output sizes must
match, it is natural to describe quantum gate operations on single qubits using 2x2 matrices, and
quantum gate operations on pairs of qubits using 4x4 matrices.

Finally, operations must preserve the normalization of the quantum state. As was explained
earlier, quantum states, and thus gates that operate on quantum states, must preserve probability.
Mathematically, this means that for an operator $\Ahat$ acting on a state $\ket{\psi}$ as $\Ahat \ket{\psi} = \ket{\psi'}$, the new state $\ket{\psi'}$ must satisfy $|\psi'|^2 = 1$. In the context of operator theory, this requirement implies that gate operations must be unitary and thus satisfy
\begin{equation}
    \Ahat \Ahat^\dagger = \Ahat \Ahat^{-1} = \Ihat.
\end{equation}
Since we regard quantum gates as square matrices, we thus require unitary matrices. Notice that
this condition is stronger than the invertibility condition, and that \textit{any} unitary operator is also
invertible. As such, typically these two requirements are just expressed as a single unitary
requirement, despite originating from different fundamental concepts.

\subsection{Single-Qubit Gates}

Building on this intuition, we can now attempt to create quantum interpretations of classical
logic gates, now represented as unitary matrices. The first gate we will attempt to reproduce is
the NOT gate. Assume we have the general single-qubit state as before
\begin{equation}
    \begin{array}{cc}
         \ket{s} = \twovector{\alpha}{\beta}, &  |\alpha|^2 + |\beta|^2 = 1.
    \end{array}
\end{equation}
Now, the NOT gate, often denoted as an operator by $\Xhat$, should transform a $\ket{0}$ state to a $\ket{1}$ state
and vice-versa, so that the amplitudes are flipped
\begin{equation}
    \Xhat \twovector{\alpha}{\beta} = \twovector{\beta}{\alpha}.
\end{equation}
It may be straightforward to see that the only matrix representation of $\Xhat$ that satisfies this is 
\begin{equation}
    \Xhat = \begin{bmatrix}
        0 & 1 \\
        1 & 0
    \end{bmatrix},
\end{equation}
which is indeed the quantum NOT gate for a single qubit.

Before we move forward with two-qubit gates, it is interesting to note that while the only nontrivial single classical bit gate is the NOT gate, this is not the case for qubits. There are many
non-trivial single-qubit gates, but two important single-qubit gates are the Z gate and the
Hadamard gate. The single qubit Z gate is given by
\begin{equation}
    \Zhat = \begin{bmatrix}
        1 & 0 \\
        0 & -1
    \end{bmatrix}.
\end{equation}
This gate simply flips the sign of the amplitude of the $\ket{1}$ state while leaving the $\ket{0}$ state
unchanged. Note that this does not change the probability of the $\ket{1}$ state, since probabilities are
squares of amplitudes; instead, it adds a phase of $\pi$.

The Hadamard gate is given by
\begin{equation}
    \Hhat = \fraconeroottwo \begin{bmatrix}
        1 & 1 \\
        1 & -1
    \end{bmatrix}.
\end{equation}
To understand the effect of this gate, consider the effect of applying $\Hhat$ to the $\ket{0}$ state:
\begin{equation}
    \begin{array}{cc}
         \ket{s} = \twovector{1}{0}, & \Hhat \ket{s} = \fraconeroottwo\twovector{1}{1} = \fraconeroottwo\big(\ket{0} + \ket{1}\big).
    \end{array}
\end{equation}
Similarly, consider the effect of applying $\Hhat$ to the $\ket{1}$ state:
\begin{equation}
    \begin{array}{cc}
         \ket{s} = \twovector{0}{1}, & \Hhat \ket{s} = \fraconeroottwo\twovector{1}{-1} = \fraconeroottwo\big(\ket{0} - \ket{1}\big).
    \end{array}
\end{equation}
These states are ‘halfway’ between the two basis states, thus the Hadamard gate generates superpositions where each basis state has a probability of 1/2. We may think of these two states as a uniform (symmetric and antisymmetric, resp.) superpositions, since the probabilities are uniformly distributed over outcomes (i.e. basis states). Note that $\Hhat^2 = \Ihat$, so applying the Hadamard gate to a uniform symmetric superposition returns the $\ket{0}$ state, and applying the Hadamard gate to a uniform antisymmetric superposition returns the $\ket{1}$ state. The Hadamard gate is typically used in circuits at the very beginning to generate a uniform superposition state that will be leveraged by the rest of the circuit.

We briefly mention a few other single qubit gates. The $\Xhat$ and $\Zhat$ gates described above are often
described as Pauli gates, of which there are three. The Pauli $\Yhat$ gate is given by
\begin{equation}
    \Yhat = \begin{bmatrix}
        0 & -i \\
        i & 0
    \end{bmatrix}.
\end{equation}
A well-known feature of Pauli operators is that the set $\{ \Ihat, \Xhat, \Yhat, \Zhat \}$ forms a basis over the space of 2x2 complex Hermitian operators. 

The phase gate $S$ is given by
\begin{equation}
    \hat{S} = \begin{bmatrix}
        1 & 0 \\
        0 & i
    \end{bmatrix},
\end{equation}
and is given this name simply because it maps a generic state $\ket{s} = \twovector{\alpha}{\beta}$ to the state $\ket{s} = \twovector{\alpha}{i\beta}$, thus creating a complex phase.

The $\pi/8$ gate is given the symbol $\hat{T}$ and is given by
\begin{equation}
    \hat{T} = \begin{bmatrix}
        1 & 0 \\
        0 & e^{i\pi/4]}
    \end{bmatrix}.
\end{equation}
Note that $\sqrt{i} = e^{i\pi/4} = \cos(\pi/4) + i\sin(\pi/4) = \frac{\sqrt{2}}{2} + i\frac{\sqrt{2}}{2}$, so the $\hat{T}$ gate is the square root of the $\hat{S}$ gate and corresponds to a $\pi/4$ rotation in the complex plane as opposed to a $\pi/2$ rotation with the $\hat{S}$ gate.

Finally, the rotation gates $\hat{R}_X(\theta), \hat{R}_Y(\theta), \hat{R}_Z(\theta)$ are given by
\begin{align}
    \hat{R}_X(\theta) &= e^{-i\theta/2 \cdot \hat{X}} = \begin{bmatrix}
        \cos \theta/2 & -i\sin \theta/2 \\
        -i \sin \theta/2 & \cos \theta/2
    \end{bmatrix}, \\
    \hat{R}_Y(\theta) &= e^{-i\theta/2 \cdot \hat{Y}} = \begin{bmatrix}
        \cos \theta/2 & -\sin \theta/2 \\
        \sin \theta/2 & \cos \theta/2
    \end{bmatrix}, \\
    \hat{R}_Z(\theta) &= e^{-i\theta/2 \cdot \hat{Z}} = \begin{bmatrix}
        e^{-i\theta/2} & 0 \\
        0 & e^{i\theta/2}
    \end{bmatrix}.
\end{align}
These gates are analogous to the rotation matrices in three Cartesian axes, and equivalently rotate
the amplitude vector on the Bloch sphere described in \cref{fig:bloch_sphere} about the corresponding axis. In
this context of rotations on the Bloch sphere, one can generalize all single-qubit gates to a
product of rotations. Namely, any arbitrary single-qubit gate $\Uhat$ can be decomposed as 
\begin{align}
    \Uhat &= e^{i\alpha}\Rhat_Z(\beta) \Rhat_Y(\gamma) \Rhat_Z(\delta) \\
    &= e^{i\alpha}\begin{bmatrix}
        e^{-i\beta/2} & 0 \\
        0 & e^{i\beta/2}
    \end{bmatrix} \begin{bmatrix}
        \cos \gamma/2 & -\sin \gamma/2 \\
        \sin \gamma/2 & \cos \gamma/2
    \end{bmatrix}\begin{bmatrix}
        e^{-i\delta/2} & 0 \\
        0 & e^{i\delta/2}
    \end{bmatrix} \\
    &= \begin{bmatrix}
        \cos \gamma/2 e^{i(\alpha-\beta/2-\delta/2} & -\sin \gamma/2 e^{i(\alpha-\beta/2+\delta/2} \\
        \sin \gamma/2 e^{i(\alpha + \beta/2 - \delta/2} & \cos \gamma/2 e^{i(\alpha+\beta/2+\delta/2}
    \end{bmatrix},
\end{align}
which equates to four parameters on three sequential one-parameter gates (and a phase scaling
parameter). This property is quite useful, and may be leveraged for the design of quantum
circuits.


\subsection{Two-Qubit Gates}
In the case of two-qubits we have four outcomes, as opposed to the two in the single-qubit case.
These are given by $\ket{00}$, $\ket{01}$, $\ket{10}$, and $\ket{11}$. Thus these states define the computational basis, and one can describe superpositions of these basis states in the general form
\begin{equation}
    \ket{s} = \alpha_{00}\ket{00} + \alpha_{01}\ket{01} + \alpha_{10}\ket{10} + \alpha_{11}\ket{11} = \fourvector{\alpha_{00}}{\alpha_{01}}{\alpha_{10}}{\alpha_{11}}
\end{equation}
Instead of measuring the entire state as in the case of a single qubit, here we may measure just
one of the two qubits, which has a ‘back action’ effect on the other. Say we measure the first
qubit. The probability of a $\ket{0}$ state on the first qubit is $|\subalpha{00}|^2 + |\subalpha{01}|^2$, and such an outcome would cause a post-measurement state of
\begin{equation}
    \ket{s'} = \frac{\subalpha{00}\ket{00} + \subalpha{01}\ket{01}}{\sqrt{|\subalpha{00}|^2 + |\subalpha{01}|^2}} = \frac{1}{\sqrt{|\subalpha{00}|^2 + |\subalpha{01}|^2}}\fourvector{\subalpha{00}}{\subalpha{01}}{0}{0}
\end{equation}

Two-qubit gates are defined by 4x4 unitary matrices. Among the most famous two-qubit gates is the CNOT gate, given by 
\begin{equation}
    \text{CNOT} = \begin{bmatrix}
        1 & 0 & 0 & 0 \\
        0 & 1 & 0 & 0 \\
        0 & 0 & 0 & 1 \\
        0 & 0 & 1 & 0 \\
    \end{bmatrix} = \begin{bmatrix}
        \Ihat & \hat{0} \\
        \hat{0} & \Xhat
    \end{bmatrix},
\end{equation}
where $\hat{0}$ is just the zero matrix. In essence, this gate flips the second qubit if the first qubit contains the $\ket{1}$ state, and does nothing otherwise. To see this, let us apply it to the $\ket{00}$ state:
\begin{equation}
    \text{CNOT}\ket{00} = \text{CNOT}\fourvector{1}{0}{0}{0} = \fourvector{1}{0}{0}{0}
\end{equation}
Thus when the first qubit is $\ket{0}$, the total state is unchanged. The same occurs to the state $\ket{01}$. Next apply CNOT to the $\ket{10}$ state:
\begin{equation}
    \text{CNOT}\ket{10} = \text{CNOT}\fourvector{0}{0}{1}{0} = \fourvector{0}{0}{0}{1} = \ket{11}.
\end{equation}
So, the second qubit is flipped when the first qubit contains the $\ket{1}$ state. The same qubit flip
occurs in the case of the $\ket{11}$ state, producing $\ket{10}$. More generally,
\begin{equation}
    \text{CNOT}\fourvector{\alpha}{\beta}{\gamma}{\delta} = \fourvector{\alpha}{\beta}{\delta}{\gamma}, \quad \forall \;\; \alpha, \beta,\gamma,\delta \in \Cb \;\;\; \text{s.t.}\;\; |\alpha|^2 + |\beta|^2 + |\gamma|^2 + |\delta|^2 = 1.
\end{equation}

There are a variety of other two-qubit gates that use the first qubit as a control, and conditionally
apply any of the single-qubit gates described above. Examples include the controlled-Z gate
(sometimes called CZ), given by
\begin{equation}
    \text{controlled-Z} = \begin{bmatrix}
        1 & 0 & 0 & 0 \\
        0 & 1 & 0 & 0 \\
        0 & 0 & 1 & 0 \\
        0 & 0 & 0 & -1 \\
    \end{bmatrix},
\end{equation}
and the controlled-phase gate (sometimes called CS), given by
\begin{equation}
    \text{controlled-phase} = \begin{bmatrix}
        1 & 0 & 0 & 0 \\
        0 & 1 & 0 & 0 \\
        0 & 0 & 1 & 0 \\
        0 & 0 & 0 & i \\
    \end{bmatrix}.
\end{equation}
Another widely used two-qubit gate is the swap gate, given by
\begin{equation}
    \text{swap} = \begin{bmatrix}
        1 & 0 & 0 & 0 \\
        0 & 0 & 1 & 0 \\
        0 & 1 & 0 & 0 \\
        0 & 0 & 0 & 1 \\
    \end{bmatrix}.
\end{equation}
This gate swaps the $\ket{01}$ state for the $\ket{10}$ state, but leaves other states unaffected. For the
generic two-qubit state, we have
\begin{equation}
    \text{swap}\fourvector{\alpha}{\beta}{\gamma}{\delta} = \fourvector{\alpha}{\gamma}{\beta}{\delta}, \quad \forall \;\; \alpha, \beta,\gamma,\delta \in \Cb \;\;\; \text{s.t.}\;\; |\alpha|^2 + |\beta|^2 + |\gamma|^2 + |\delta|^2 = 1.
\end{equation}


\subsection{The Quantum Universal Gate Set}

One can quickly see that the richness of the quantum description yields many more usable
operations in comparison to the classical case; one has a variety of non-trivial single-qubit gates
in contrast to only one non-trivial single bit gate. The basic intuition behind this feature is that
quantum gates can cause complex amplitudes to \textit{interfere} with each other, thus canceling out
quantum amplitudes. The feature of having a much larger (uncountably infinite) gate set is much
more pronounced with larger multi-qubit operators. The trade-off is that there is significant
added complexity to define the universal gate basis set, that is, the set of quantum gates that can
produce any unitary operator with sufficient precision. 

The challenge with a universal quantum gate set is that it asks to compose a finite set of
operators that when put in sequence can produce an uncountably infinite set of n-qubit operators.
Thus, the claim of classical universality is weakened for the quantum case by only requiring that
we produce any unitary operator with \textit{sufficient precision}. Some requirements for a universal
quantum gate set are that it must be able to \textit{create} superposition (e.g. Hadamard), that it must be
able to \textit{create} entanglement, and that it must be able to create complex amplitudes as well as real
ones.

It has been shown \cite{aaronson2022introduction} that one such universal quantum gate set is the set $\{\text{CNOT}, \hat{S}, \Rhat_X(\pi/4)\}$. This set is not unique; one can substitute the $\Rhat_X(\pi/4)$ gate for nearly any rotation gate. Such a universal set enables the quantum computing paradigm to recreate, and generalize the classical
computing paradigm.


\subsection{Fundamental Challenges of Quantum Computing}

Here we describe one of the major difficulties in realizing quantum computers. The increased
richness in the expressibility of a quantum superposition state also leads to one of the hardest
problems in realizing quantum computing hardware, the problem of quantum error correction
\cite{nielsen2010quantum}. To elucidate this, consider first the classical computing case. 

There is always noise that can induce errors in any information processing channel. As a result,
even in classical computing there can sometimes be random errors that change a bit from a 0 to a
1, which are called bit flip errors. In classical computing, any channel of communicating
information typically appends redundant information to a binary string in order that one can use
the redundant information to decode the binary string and recover the intended information
despite the error. For example a repetition code with majority voting is a simple classical error
correction scheme, where if you intend to communicate a 0 bit, you instead send 000, and
similarly to send the 1 bit you send 111. At the receiving end, you simply decode the bit of
information based on the majority of 0 or 1 bits communicated, such that you can always protect
against a single bit flip error.

In the quantum error correction case, since our qubit states are superpositions of basis states,
errors can be understood as changing the probability amplitudes of outcomes. Consider a single
qubit as before:
\begin{equation}
    \begin{array}{cc}
         \ket{s} = \alpha \ket{0} + \beta \ket{1} = \twovector{\alpha}{\beta}, & |\alpha|^2 + |\beta|^2 = 1.
    \end{array}
\end{equation}
As in the case of the $\Xhat$ gate, the qubit flip yields:
\begin{equation}
    \ket{s'} = \twovector{\beta}{\alpha} = \beta\ket{0} + \alpha\ket{1}.
\end{equation}
We can similarly apply the classical repetition code for qubits as:
\begin{align}
    \ket{0} &\rightarrow \ket{0}_L \equiv \ket{000}, & 
    \ket{1} &\rightarrow \ket{1}_L \equiv \ket{111}
\end{align}
where $\ket{\cdot}_L$ corresponds to a logical qubit, so that the single qubit state becomes a three-qubit state:
\begin{equation}
    \ket{s} = \alpha \ket{0}_L + \beta \ket{1}_L.
\end{equation}
As before, we can diagnose and correct a single qubit flip by majority voting.

When viewed as a Pauli rotation on the Bloch sphere, a single bit flip is equivalent to a $\pi$ rotation
about the x-axis. A fundamental challenge arises upon realizing that the error may rotate our
qubit not only about the x-axis, but about \textit{any} axis in 3D. Thus for a single qubit, we must also
perform a similar error diagnosis on the \textit{phase} flip of a qubit, which has a similar flavor to the bit
flip diagnosis, but in a rotated basis corresponding to the x-axis on the Bloch sphere
\begin{align}
    \ket{+} &= \fraconeroottwo \big( \ket{0} + \ket{1}) \\
    \ket{-} &= \fraconeroottwo \big( \ket{0} - \ket{1}) \\
\end{align}
Logical qubits can be formed in this basis such that a phase flip is diagnosed using a similar
majority voting repetition code

A famous result in quantum error correction is known as the Shor code, which can correct an
\textit{arbitrary} error (i.e. any single qubit rotation error), by encoding a 9-qubit logical qubit, however
there are codes that can correct any single qubit error by encoding as small as a 5-qubit logical
qubit. Thus a quantum computer with only single qubit errors must implement 5 times the
number of qubits to attain a given number of logical qubits. 

Coupled to this issue is that circuits of increasing numbers of qubits suffer from \textit{cross-talk} errors,
which is when qubits in different channels become undesirably entangled to each other, which
requires another level of quantum error correction. There are a variety of quantum errors that
appear in hardware that must be diagnosed and corrected, which leads to a complex problem of
quantum error correction over arbitrary width quantum circuits. A major challenge in quantum
computing is finding codes that reduce the necessary number of qubits to achieve a logical qubit,
and building a large enough quantum computer with enough fault tolerance to be able to perform
useful operations on these logical qubits.


\section{Introduction to Classical Deep Learning Algorithms}

The primary goal of machine learning and more specifically deep learning is to optimally
approximate some functional mapping that encodes a challenging task. These tasks historically
drew heavy inspiration from the everyday tasks that are accomplished by the animal brain.
Today, a vast variety of interesting problems are addressed using machine learning and deep
learning, from understanding protein folding for better drug discovery \cite{jumper2021highly}, to predicting
complex weather patterns \cite{pathak2022fourcastnet}, to processing and/or translating language \cite{openai2023gpt}, to autonomous
driving \cite{lu2023imitation}, to realizing nuclear fusion technology for green energy production \cite{degrave2022magnetic}.

For example when a fox sees the movement of an animal in the distance, its visual cortex must
quickly determine if that animal is a bear and the fox should scurry away, or if the animal is a
rabbit and the fox should pursue it. This general perception task is highly studied in machine
learning literature and known as the task of classification. Here, the natural processes of the brain
that classify the visual image of an animal as a bear or a rabbit can be understood as a functional
mapping between images and classes of animals.

Another example is based on the motor cortex, where a human might want to pick up a glass of
water in order to drink from it. The motor cortex evaluates the current position, velocity, and
acceleration of the arm, and the current position of the cup, and must determine the electrical
signals that are sent to the arm to cause it to extend and grasp the cup. This general task is also
highly studied in robotics literature and known as the task of control (i.e. controlling the hand
and arm to grasp the cup). Here again, the natural processes of the brain can be understood as a
functional mapping between generalized locations and muscular actuation signals.

This sort of input-output mapping representation of a process or relationship is extremely
flexible and general, perhaps universal. Using this framework, the goal of machine learning is to
closely approximate the inherent relationship between input and output by a function that
contains many, often millions, of flexible parameters often referred to as weights. The mapping is
most commonly represented, or modeled, using a highly parameterized Artificial Neural
Network (ANN), and the learning in machine learning refers to optimizing the parameters of the
model so that it can mimic the process or relationship to high accuracy and precision. The
adaptability and expressibility of ANNs have been one of the biggest motivators of their wide
spread use. In this section, the basics of ANNs will be covered.


\subsection{Overview of Classical Machine Learning Problems and Terminology}

In machine learning the general goal is to learn a model from a system with known inputs, and
sometimes outputs, so that the model can predict or enhance understanding of the underlying
data. In the most general case there are two types of machine learning algorithms; supervised and
unsupervised. In the unsupervised case, only the inputs to a system are known and the goal is
generally to better describe the data itself. For example, a common unsupervised learning task is
called clustering, wherein the machine learning model is attempting to find patterns in the data
that identify common characteristics of portions of the data so that homogeneous data is grouped,
or clustered, together. In the supervised case, both the inputs and outputs of the system are
known and the goal is to accurately predict the systems outputs given the inputs. We will focus
on the supervised case as that is more related to our current work.

In the supervised machine learning framework there are again two tasks that machine learning
performs; classification and regression. In the classification task machine learning models seek to identify membership of the input data to categories known as classes. For example,
determining if an image contains a cat or a dog is understood as determining if the object in the
image belongs to the ‘cat’ class or the ‘dog’ class. This type of model will often be referred to as
a classifier, as its purpose is generally to assign the correct class to an input, where the outputs
are effectively binary ‘yes or ‘no’ for class membership of each class. The regression task is very
similar except that instead of being binary in the output, models generally output a continuous
value, e.g. given an image of a dog, predict the weight in kilograms of said dog. The differences
in these tasks usually comes down to the type of data available and the setup of the optimization
problem.

Once the task and model have been defined, the model’s parameters are optimized such that the
model fits the data as closely as possible. This optimization procedure is often referred to as
training, and is the critical phase of a machine learning algorithm. Here, some function $\calF$ that quantifies some abstract measure of distance, or error, between the predicted labels $\yhat$ and the true labels $y$ of the training data $x$ is minimized as
\begin{equation}
    \theta_{\text{best}} = \argmin_\theta \calF \big(\yhat(\theta,x), y\big),
\end{equation}
where $\calF \big(\yhat(\theta,x), y\big)$ is the e function to be minimized (e.g. mean squared error) measuring some notion of ‘badness’ of the model’s prediction of the class label. This function is sometimes called
an objective function, a cost function, or a loss function depending on the community. All machine learning models can be represented by an equation that maps the input to the desired output, and in this case the predicted labels $\yhat$ are a function of the model parameters $\theta$ and the training data $x$.

Before beginning to train, the available data is usually divided into three categories: training
data, validation data, and test data. The training data is the data that is used to optimize the model
parameters, while the validation data is used to prevent overfitting. Overfitting occurs when the
optimization process learns the training data too well and does not generalize to other data.
During training, the validation data is evaluated using the same function as the training data,
generally referred to as the validation loss, however the model is never updated using the
validation loss. The validation loss is monitored over training iterations, and overfitting is
generally indicated by an increase in the validation loss for a decreasing training loss. When the
validation loss begins increasing, training can be stopped to prevent overfitting. The final step in
the machine learning process is the testing phase. Here the model is evaluated to determine true
performance. This phase requires a completely blind set of data called test data, i.e. neither the
validation nor training data can be used. This ensures that the performance of the model on the
test data captures what typical performance would look like in practice.

If we consider a general classification task such as the one described above, the goal of training
is to separate the input space into multiple regions where each region contains only a single class
label. This clearly requires that the data is separable in some way; and the simplest case is where
the data is linearly separable. Let’s assume that our data has two classes, if a straight line can be
drawn between the classes that data is linearly separable and any machine learning model will be
able to correctly classify this data. This very rarely happens in real world data, thus many
advanced models have been developed to handle data of various complexities, such as data that
is not linearly separable. One such model will be covered next.


\subsection{Neural Networks}

The beginnings of ANNs trace back to 1957 with Rosenblatt's perceptron \cite{rosenblatt1957perceptron}. The perceptron classifier is a simple linear classifier/regression model. In the perceptron a group of weights are used to transform the input variables into the desired output. The perceptron uses the equation, $\vy = \vW \vx$, where $\vx$ is the input vector, $\vW$ are the weights, and $\vy$ is the output which is either a scalar value or a new vector. This is a linear mapping which is severely limited in its real world application as demonstrated by the following example.

\begin{figure}[t]
    \centering
    \includegraphics[width=0.95\textwidth]{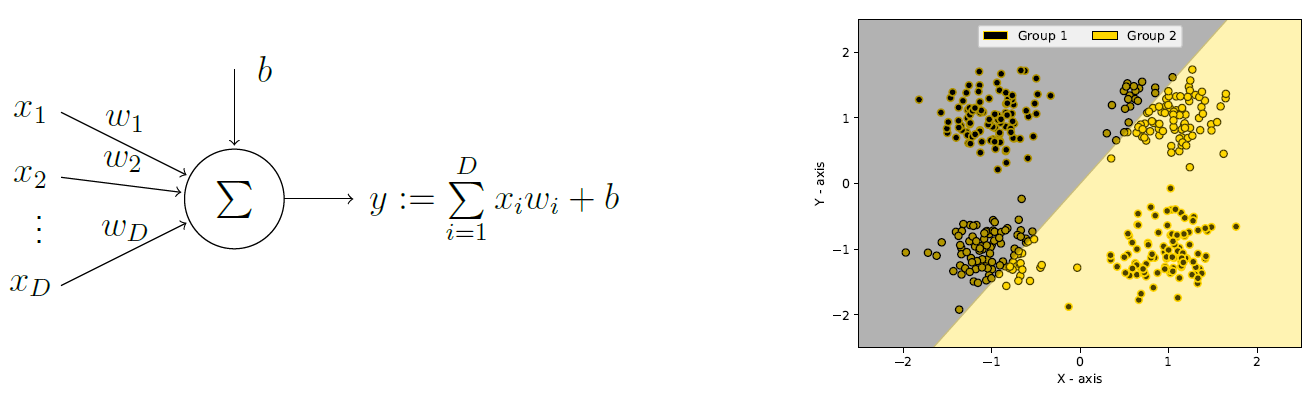}
    \caption{Perceptron Example}
    \label{fig:perceptron}
\end{figure}

The XOR problem is often used for illustrations on the decision boundaries obtained with neural networks. The XOR problem generally consists of four clusters of data which are grouped into two classes. The clusters are generally created in a square, i.e. the four cluster centers are located at (-1, -1), (-1, 1), (1, 1), and (1, -1), with the diagonal clusters labeled as one class and the off-diagonal clusters as the opposite class. The perceptron is shown graphically on the left side of \cref{fig:perceptron}, and the image on the right shows the limitations of the perceptron in solving the XOR problem. In the image, the different colors each indicate a different class. Since the perceptron is a linear classifier it can only define a simple linear decision boundary. Therefore, the perceptron is unable to correctly separate the diagonal and off-diagonal classes.

\begin{figure}[t]
    \centering
    \includegraphics[width=0.95\textwidth]{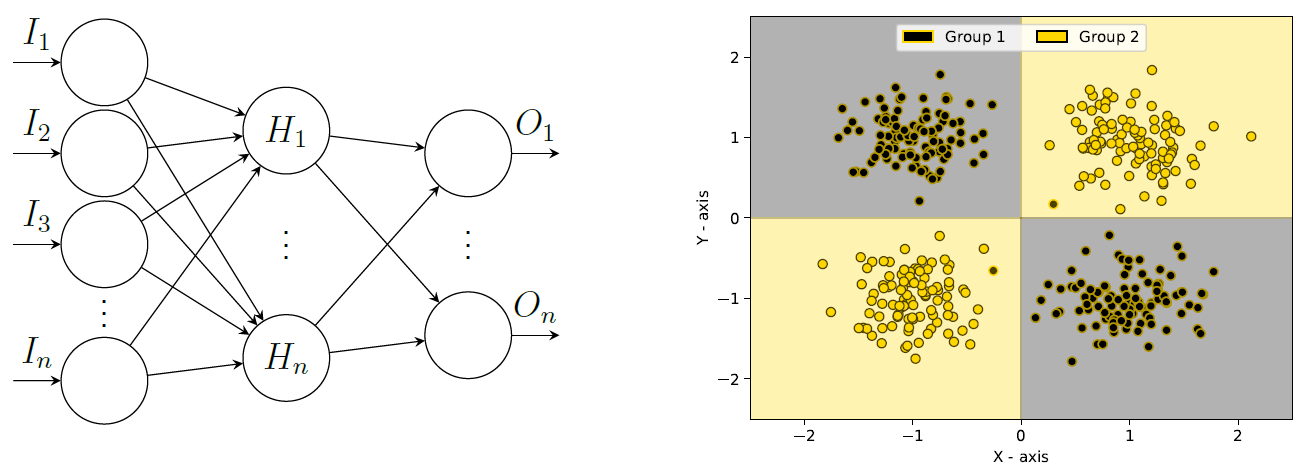}
    \caption{Multilayer Perceptron Example}
    \label{fig:multilayer_perceptron}
\end{figure}

To improve upon the perceptron it was found that performance could be greatly improved by combining multiple perceptrons together. The combination process involves creating layers of perceptrons. A layer is generated by having several perceptrons in parallel, with each perceptron using the same inputs. When combined in this fashion each perceptron is generally called a neuron. As the single perceptron case is simply an inner product between the inputs and the weights, a layer can be seen as a linear mapping from the input space to the space defined by the weights of each neuron. Thus each layer is computed as a matrix multiplication. The output from each linear mapping is then fed into a non-linear function often referred to as an activation function, which is named after its approximation of the step activation of a biological neuron. This process is often referred to as the Wiener method \cite{wiener1966nonlinear}, which has the non-linear activation following the summation in \cref{fig:perceptron}. The purpose of the non-linear activation is to increase expressibility of the network, as without non-linearities a series of linear mappings will always reduce to a single linear mapping. Stacking multiple layers together, by connecting the outputs of a layer to the inputs of the following layer, creates what is called the Multi-Layered Perceptron (MLP) \cite{rosenblatt1962principles}, and is shown in the left side of \cref{fig:multilayer_perceptron}. In the diagram each circle is considered a neuron. The outputs from the final layer, generally called the logits, are often converted to probabilities via the softmax function, which is defined as
\begin{equation}
    \sigma(\vz)_i = \frac{e^{\beta z_i}}{\sum_{j=1}^K e^{\beta z_j}}, \;\;\; i = 1, \dots, K,
\end{equation}
where $K$ is e number of logits in the output layer, $\vz$ is the vector of output logits with $z_i$ being
the $i^{th}$ element of the vector, and $\beta$ is referred to as the temperature which is a hyperparameter
that controls the smoothness of the softmax function.

With the addition of multiple layers and non-linearities, the MLP is now capable of classifying data that is not linearly separated, as in the right side of \cref{fig:multilayer_perceptron}. The stacking of layers allows the MLP to define multiple regions that can be separated linearly by the final classification layer. For example, in the XOR problem shown in the right side of \cref{fig:multilayer_perceptron} the hidden layer can divide the feature space into the four quadrants shown. The classification layer then classifies the quadrants into their proper classes as the first layer projects the data into a space where the quadrants are linearly separable. To be considered a MLP there must be at least three layers: the input layer, the output layer, then at least one hidden layer. The hidden layer(s) fall between the other two layers as shown in the left side of \cref{fig:multilayer_perceptron}. In general multi-layer networks, there are many layers of matrix-vector multiplications which can be expressed as 
\begin{equation}\label{eq:multi-layer}
    f(\vx_n; \vW) = \sigma_{N_h} \Bigg(\vW_{N_h}\sigma_{N_{h-1}}\bigg(\vW_{N_{h-1}}\sigma_{N_{h-2}}\Big(\cdots\sigma_1(\vW_q\vx_n)\Big)\bigg) \Bigg),
\end{equation}
where $\vx_n$ is an input, $N_h$ is the number of hidden layers, $\sigma_h(\cdot)$, $h = 1, 2, \dots , N_h$ are activation functions, and $\vW$ are the weights. To describe the potential capabilities of the MLP the Universal
Approximation Theorem was proved in \cite{hornik1989multilayer} which states that MLPs are capable of approximating any continuous function to an arbitrary accuracy given that the hidden layer is of sufficient size.


\subsection{Shared Weight Neural Networks}

\begin{figure}[t]
    \centering
    \includegraphics[width=0.75\textwidth]{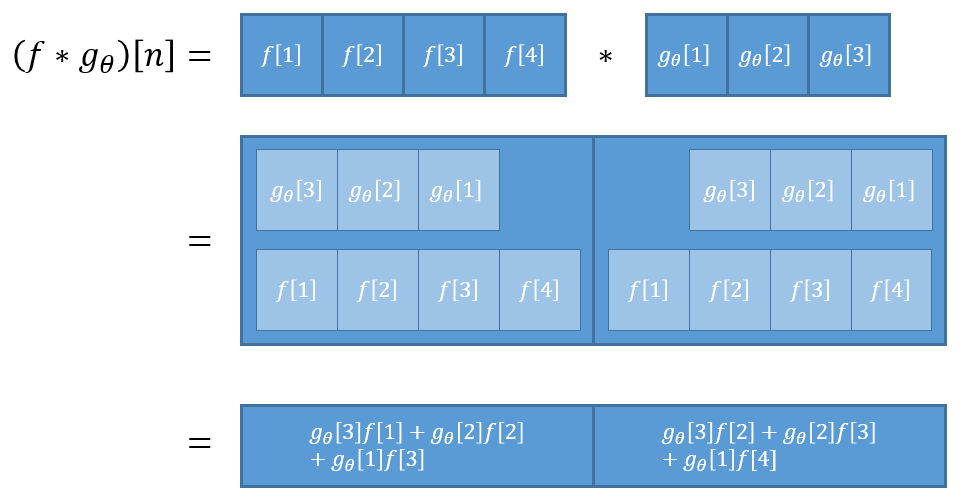}
    \caption{Simple convolution example}
    \label{fig:conv}
\end{figure}

The standard MLP is effective for a broad category of tasks, however the MLP also introduces a bias toward interconnectivity of every data point. While this is often a good strategy, it can introduce significant redundancy, and specifically many classification problems rely on images, for which local relationships are more important. In adding more capability to the MLP, the next big advancement was the shared weight network from \cite{rumelhart1985learning}. This is easiest to visualize via the convolutional neural network (CNN) \cite{lecun1989backpropagation}. Convolution networks are based on the standard convolution operation defined as:
\begin{equation}
    (f \ast g_\theta)[n] = \sum_{m=0}^{N-1}f[m]g_\theta[n-m],
\end{equation}
where $f$ and $g_\theta$ are functions to be convolved, and $g_\theta$ is parameterized by $\theta$ and is often referred to as the convolution filter. Note that convolution is usually implemented as correlation in convolution networks, because correlation requires fewer operations and the convolution filter is learned therefore the two are effectively equivalent. To better explain what is happening see \cref{fig:conv}. In the figure a simple convolution example is shown using two vectors, one of size 4 and another of size 3.

For this demo only the valid portion of the convolution is used meaning only the locations where
two vectors fully overlap are used, these two positions are shown in the center line of the figure.
The last line of the diagram shows the output of the convolution as an equation of the individual
elements of $f$ and $g_\theta$. The last line can also be represented as a matrix multiplication as in:
\begin{equation}
    (f \ast g_\theta)[n] = \begin{bmatrix}
        g_\theta[3] & g_\theta[2] & g_\theta[1] & 0 \\
        0 & g_\theta[3] & g_\theta[2] & g_\theta[1]
    \end{bmatrix}
    \fourvector{f[1]}{f[2]}{f[3]}{f[4]}.
\end{equation}

\begin{figure}[t]
    \centering
    \includegraphics[width=0.25\textwidth]{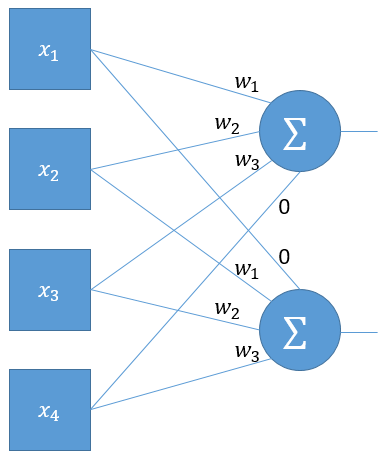}
    \caption{Simple convolution layer example}
    \label{fig:conv_layer}
\end{figure}

With this representation if we write the 2 × 4 matrix as $\vW$ with $w_n = g_\theta[n]$, and the vector of $f[n]$ as $\vx$, with $\vx_n = f[n]$, we are left with exactly the equation of a perceptron, $\vy = \vW \vx$ as was
seen before. With this new representation we can rewrite the convolution operation we started with as the simple single layer neural network shown in \cref{fig:conv_layer}.

In this representation the weights $\{ w_1, w_2, w_3 \}$ are shared across both of the neurons represented by the summation symbols, thus a small set of weights can be shared across the input space. This type of network is in general applied to images, and can dramatically reduce the number of network parameters, which can reduce overfitting, reduce training time, and reduce model complexity.


\subsection{Gradient Descent}

\begin{figure}[t]
    \centering
    \includegraphics[width=0.35\textwidth]{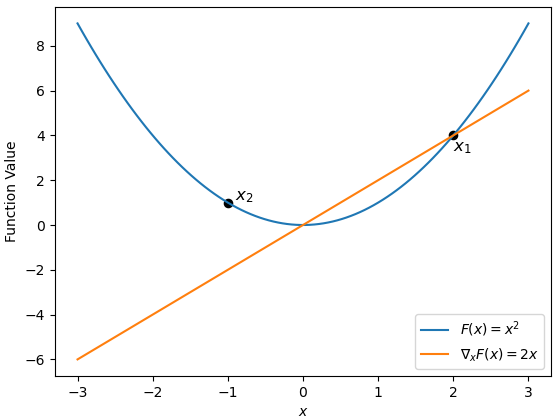}
    \caption{Example of a gradient calculation}
    \label{fig:grad}
\end{figure}

Training for all types of neural networks uses some flavor of gradient descent. Gradient descent
is an optimization strategy that is widely used for fitting many different types of models and
data. Assuming a convex function, for a given point the sign of the gradient points away from the
minimum. For example, in \cref{fig:grad}, assume we are trying to find the minimum of the function $F(x)$ starting at the point $x_1 = 2$. Computing the gradient of $F(x)$ at $x_1$ results in $F'(x_1) = 4$. If $F'(x_1)$ is added to $x_1$ we would move in the opposite direction of the actual minimum located at $x=0$. This also happens for $x_2$ on the opposite side of the minimum. With this example we can see that by taking a small step in the direction of the negative gradient we can move closer to the true minimum function value.

Gradient descent is simply repeatedly over many iterations (update evaluations) computing the
gradient and taking a small step in the direction of the gradient. Therefore, the rule for updating
any model via gradient descent is as follows,
\begin{equation}\label{eq:grad}
    \vx_{\text{new}} = \vx_{\text{old}} - \lambda \nabla_x F(x),
\end{equation}
where $F(x)$ is the objective function parameterized by the parameter $\vx$ and $\lambda$ is the step size that
controls how far to move in the direction of the gradient. Repeatedly applying \cref{eq:grad} until convergence will return the value of $\vx$ that minimizes $F(x)$.

Gradient descent for machine learning follows the same basic formula as above except that the
function being minimized typically takes the form,
\begin{equation}
    \calF(\vW, \vX, \vY) = \frac{1}{n}\sum_{i=1}^n \calL(\vW, \vx_i, \vy_i),
\end{equation}
where $\calL(\vW, \vx_i, \vy_i)$ is a function often called the loss function (squared error for example), which
depends on the model parameters $\vw$, training data $\vx$, and training labels $\vy$. The loss function is
then summed over the $n$ samples available in the training dataset. This is an unbiased estimator
for the expected value of $\calL(\vw, \vx_i,\vy_i)$ over the inputs $\vx_i$. For example, using the perceptron
discussed earlier as the model with mean squared error as the loss function yields,
\begin{equation}
    \calF(\vw,\vX,\vy) = \frac{1}{2n}\sum_{i=1}^n (y_i-\vw\T x_i)^2.
\end{equation}

Here the training data $\vX$ and the training labels $\vy$ are known, therefore optimization is done
solely on the model parameters $\vw$ Evaluating the gradient and using Equation \cref{eq:grad} generates
the following update equation for the perceptron algorithm,
\begin{equation}
    \vw_\text{new} = \vw_\text{old} - \frac{\lambda}{n}\left(\sum_{i=1}^n \big(y_i - w_\text{old}\T \vx_i \big)\right)\vx_i.
\end{equation}
This method is commonly referred to as the batch gradient descent update. Here the entirety of
the training set is used to compute a single update to the model, so there is only a single update
in each iteration. 

Another version of gradient descent is commonly called stochastic gradient descent. In this
variant instead of using the entire training dataset to compute a single gradient update, only a
single data point is used for each update. Therefore, the stochastic gradient descent update
equation is reduced to
\begin{equation}
    \vw_\text{new} = \vw_\text{old} - \lambda \big(y_i - w_\text{old}\T \vx_i \big)\vx_i.
\end{equation}
The biggest difference between the two methods is in how many updates are performed. For the
batch gradient descent there was only one update for each pass of the dataset, however, for the
stochastic gradient descent method there will be updates each pass. A pass through the dataset
is commonly called an epoch. A way to interpret the differences between the batch and the
stochastic versions is that, effectively, the batch method computes the average of all the
individual updates to compute its one update. This allows a smoother convergence to the correct
solution. The stochastic update leads to a much noisier convergence curve as the model is
reacting to each data sample individually. The noisy convergence curve cab be beneficial though,
as the randomness in the updates allows the model to jump out of local minima during training to
potentially find better solutions.

There is a third type of gradient descent that is far more widely used, especially in the deep
learning community, which is the mini-batch method. Mini-batch gradient descent is a hybrid of
the batch and stochastic methods. Whereas the batch gradient method uses the entire dataset to
compute the gradient, the mini-batch method uses only a small subset, larger than 1, to compute
the gradient. This also allows for updating the model multiple times during an epoch but does not
require updating after every single training sample. It also incorporates stochasticity by randomly
sampling the mini-batches of data. The full algorithm for the mini-batch gradient descent is
shown in \cref{alg:sgd} below. This method incorporates the benefits of both methods by allowing
the optimizer to have some randomness in the updates like the stochastic method but limits the
amount of randomness by controlling the batch size. A small batch size increases the
randomness, and thus optimization behaves more like the true stochastic version, while
increasing the batch size reduces the randomness so the optimization behaves like the standard
batch gradient descent method.

\begin{algorithm}
\caption{Mini-batch gradient descent algorithm}\label{alg:sgd}
\begin{algorithmic}
\Ensure Training data $\vX$, training labels $\vy$, number of epochs $N$, batch size $m$
\State Randomly initialize model parameters $\vw$
\For{$i=1,\dots,N$}
    \State Randomly shuffle $\vX$ and $\vy$
    \State Divide $\vX$ and $\vy$ into batches of size $m$, save number of batches $M$
    \For{$j=1,\dots,M$}
        \State $\vw \gets \vw - \lambda \frac{1}{2m}\sum_j^m \nabla_\vw \calL(\vw,\vX_i,\vy_i)$
    \EndFor
\EndFor
\State \Return $\vw$
\end{algorithmic}
\end{algorithm}

\subsection{Backpropagation: Gradient Descent for Neural Networks}

Gradient descent as it is defined above works well in many cases, however with neural networks the sheer number of parameters and serial operations can make differentiating with respect to each parameter inefficient. Combining Equation \cref{eq:multi-layer} with the mean squared loss (for simplicity) leaves:
\begin{equation}
    f(\vx_n, y_n, \vW) = \half \Bigg( y_n - \sigma_{N_h} \bigg(\vW_{N_h}\sigma_{N_{h-1}}\Big(\vW_{N_{h-1}}\sigma_{N_{h-2}}\big(\cdots\sigma_1(\vW_q\vx_n)\big)\Big)\bigg) \Bigg)^2,
\end{equation}
as the full loss function to be optimized. Following the process described above for gradient
descent we would differentiate with respect to $\vW$ to find the update equation. However, a single
expression cannot be found to optimize over each individual $\vW_{N_h}$ simultaneously as was done
before. A separate update function could be found by differentiating with respect to each $\vW_{N_h}$,
however as mentioned this is inefficient as each set of parameters will have a unique update
equation, also many of the calculations for each of these differentiations are repeated. On top of
that this process is not universal for all neural networks. Backpropagation is a way to compute
the gradients in a systematic fashion to efficiently calculate all the gradients in a neural network
one layer at a time that can be universally applied to all neural networks that also minimizes the
amount of duplicate calculations. At a high level backpropagation can be thought of as a large
chain rule. The per-layer loss gradient, often called the local gradient, is computed backwards
across layers of the network. In this manner the local gradient for layer $i$ is computed with
respect to only the inputs and outputs of layer $i$. When applied in a chain-rule like manner the
loss is passed backwards, starting at the output, through each layer. Each layers’ parameters are
updated in accordance with how much those parameters attribute, via the gradient, to the total
loss. For a more detailed explanation see \cite{bishop1995neural}.


\section{Current State of Quantum Computing}


\subsection{Quantum Circuit Architecture}

The currently dominant approach to quantum computing is to create a quantum circuit. This
approach is similar to classical digital logic circuits in organization and structure, but there are
some key differences. In classical digital logic circuits, a set of bits are typically initialized in the
binary 0 state, and are fed through logic gate operations sequentially until the computation is
complete and the results are read out. These circuits are often represented graphically by
sequences of lines between symbols representing logic gates that eventually lead to an output.
Critically, the values of the bits of the circuit can be measured at any point throughout the circuit
without affecting the rest of the circuit. 

With quantum circuits, qubits are similarly prepared in some initial state, usually the qubit’s zero
state, and are also fed through sequences of gate operations that are also graphically represented
by symbols (typically rectangles) connected by lines that eventually lead to some output which is
read by quantum measurement. However a major difference from classical digital circuits is that
a qubit measured before the end of the circuit will have significant effects on the rest of the
circuit. This is because there is an associated back-action as a result of any quantum
measurement, and often the measurement back-action collapses the quantum wave function of
the measured qubit, reducing it to a single classical value from that point on. This aspect is
represented graphically in quantum circuits using double lines for classical values and single
lines for quantum values. 

\begin{figure}[t]
    \centering
    \includegraphics[width=0.5\textwidth]{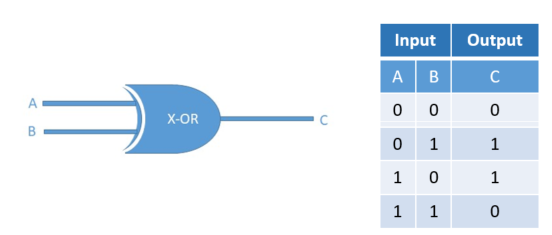}
    \caption{The XOR gate.}
    \label{fig:xor_gate}
\end{figure}

\begin{figure}[t]
    \centering
    \includegraphics[width=0.5\textwidth]{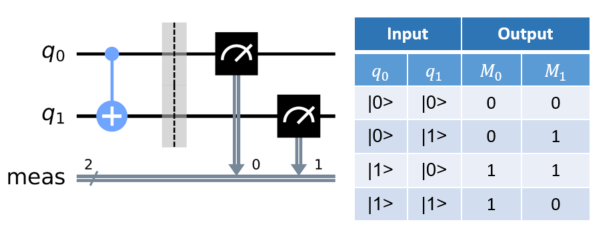}
    \caption{Simple quantum CNOT gate.}
    \label{fig:quantum_CNOT}
\end{figure}

Another major difference from digital logic is regarding circuit structure. As described in the
fundamentals of quantum computing section, digital logic gates are allowed to have a different
number of inputs than outputs, while quantum gates must be unitary and thus have equal
numbers of inputs and outputs. For example classical gates such as AND, OR, and XOR gates
have two inputs and only one output, such that the operations are irreversible and the total
number of bits at any given point in the circuit is not fixed. Since quantum gate operations must
be unitary and reversible, the total number of qubits is conserved throughout the circuit\footnote{Note that quantum measurement performed before the end of the circuit may often be graphically represented as reducing the number of qubits, however these qubits continue to exist classically after measurement}. For
example, consider the classical XOR digital logic gate depicted in \cref{fig:xor_gate} and the quantum CNOT gate depicted in \cref{fig:quantum_CNOT}. These two
gates have similar outputs that produce similar truth tables. For the digital XOR gate shown
in \cref{fig:xor_gate}, two inputs, A and B, are fed in and one output, C, is produced. If A and B are the same,
the output is a value of 0. If A and B are different, the output is a value of 1. 

In contrast, the quantum CNOT gate uses the state of one of the input qubits as a control qubit,
and determines the action on the other qubit, the target qubit, based on the control qubit’s state.
This is represented graphically in \cref{fig:quantum_CNOT}, where $q_0$ is the control qubit and $q_1$ is the target qubit. The CNOT gate itself is represented by a unique symbol. The $\oplus$ symbol represents the
NOT operation being applied to $q_1$, which is connected to the $q_0$ qubit and terminates in a dot
representing $q_0$ as the control of the NOT operation. If $q_0$ is in the zero state, $q_1$ is unaffected,
while if $q_0$ is in the one state, the NOT operation will be applied to $q_1$ and its state will be
flipped, which is equivalent to rotation by $\pi$ about the x-axis. In contrast to the digital XOR gate,
both input qubits are conserved throughout the calculation and are measured at the end of the
circuit.

Also unique to quantum computing is that the output of the quantum measurement process is not
the quantum state. Instead, an observable associated with the qubit is measured, yielding one of
the possible eigenvalues of the operator corresponding to the the quantum state the qubit was in.
This is highlighted in the truth table in \cref{fig:quantum_CNOT}, where the inputs are quantum states
represented in ket notation, and the outputs are the measured eigenvalues of the operator, and
associated with the two possible quantum state outcomes.

\begin{figure}[t]
    \centering
    \includegraphics[width=0.5\textwidth]{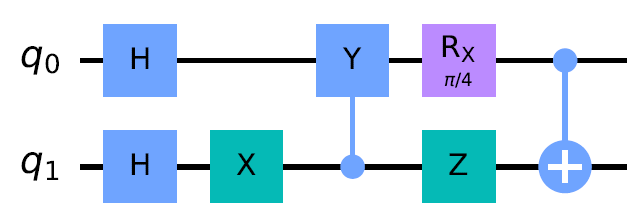}
    \caption{Simple quantum circuit example}
    \label{fig:qc_example}
\end{figure}

In \cref{fig:quantum_CNOT}, $M_0$ is the measurement outcome on $q_0$, and $M_1$ is the measurement outcome on
$q_1$. If we omit the $M_0$ column, then we recover the truth table for the classical XOR gate. Note
however that this truth table does not include the continuum of possible superpositions of qubit
states, which are valid inputs in the analogous quantum gate. Additionally, the $q_1$ qubit is still
present at the end of the circuit which, for unitary gates, allows for reversibility and
reconstruction of the input states given the output and operation applied. Graphically, gates are
applied sequentially from left to right, as depicted in \cref{fig:qc_example}. 

There are some gates that operate on larger numbers of qubits. They can be generically
represented graphically by rectangles that cover multiple qubit lines, however some specific
multi-qubit gates have their own representations. Any unitary single qubit operation can be
turned into a controlled operation that depends on the state of another qubit. This is shown by the
appropriate box/symbol on the qubit to be operated on, with a vertical line extending from the
box to the horizontal line of the controlling qubit with a dot placed at their intersection. 

Another contrast to classical logic circuits is that in quantum circuits, one measurement is not
enough to deduce the quantum state of the output qubit(s). For example, a qubit in the
superposition state 
\begin{equation}
    \ket{q} = \fraconeroottwo\ket{0} + \fraconeroottwo \ket{1} = \twovector{\fraconeroottwo}{\fraconeroottwo}
\end{equation}
has a measurement outcome of 0 with probability $1/2$ and a measurement outcome of 1 with probability $1/2$. Thus several measurements of the same qubit must be made in order to deduce hat it is in the given superposition state. Thus the expectation value is estimated by repeating the measurement many times. Also, since the measurement value only returns the magnitude, the expectation value will be equivalent for a set of quantum states that have the same magnitude but different phase, for example the state
\begin{equation}
    \ket{q} = \fraconeroottwo\ket{0} - \fraconeroottwo \ket{1} = \twovector{\fraconeroottwo}{-\fraconeroottwo}
\end{equation}
These two states are identical except for a phase factor, and this type of measurement protocol
(with a single Z measurement) does not have the resolution to discern between such states. There
are workarounds, for example measuring with respect to the x-axis instead of the z-axis, but this
leads to extra care that is necessary in designing quantum circuits. For further reading on the
fundamentals quantum gates and quantum information, see references \cite{nielsen2010quantum,kaye2006introduction}. 

\subsection{Embedding Classical Data in Quantum Circuits}

Since quantum data and classical data are inherently different in nature, methods must be used to
encode classical data in a way that is usable in quantum circuits. Currently, there are two main
strategies for building quantum machine learning circuits that use classical data. The first
strategy is to use classical dimensionality reduction techniques to reduce the dimension of the
classical data to match the number of qubits available in the circuit, such as principle component
analysis. In order to embed binary data specifically, an additional step is necessary to convert the
reduced dimension data to binary values. An example of a method to reduce dimensionality and
convert to binary values is shown in \cref{sec:dim_reduction}.

Another type of encoding is often called gate encoding. In this paradigm the original floating
point data is encoded directly into a quantum circuit with the use of rotation gates. For this type
of encoding, the original data is normalized to be in the range of $[0, \pi]$. This range is used to
ensure that large and small values are not unintentionally confused for being close together, as
they could be if the full $[0, 2\pi)$ range was used. In cases where the data has fewer or the same
dimensions as the number of qubits, each value of the original data can be directly encoded into
the rotation parameter of a rotation gate. In this setup each dimension of the data is encoded by
exactly one rotation gate per qubit during the encoding, which can then be used by additional
circuit elements for machine learning.

An extension of this method, called block encoding, takes this method and applies it to higher
dimensional data. Here the data is represented as a quantum circuit containing many rotation
gates applied to the same qubits in order to generate a unique encoding for each data point. Four
numbers are important for the design of this encoding scheme: the dimensionality of the data
($D$), the number of qubits ($Q$), the number of layers of the circuit ($L$), and the number of gates
per block ($G$). The values of $D$ and $Q$ should already be known and the values of $L$ and $G$ are
design variables. To select the values of $L$ and $G$ follow the rule that $D\leq QLG$ while also trying
to minimize the product $QLG$. For example, if the data consists of 192 total dimensions, and we are using a quantum computer/simulator with 16 qubits, we can set $L$ to 4 and $G$ to 3. With the design settled, the circuit can be created

\begin{figure}[t]
    \centering
    \includegraphics[width=0.95\textwidth]{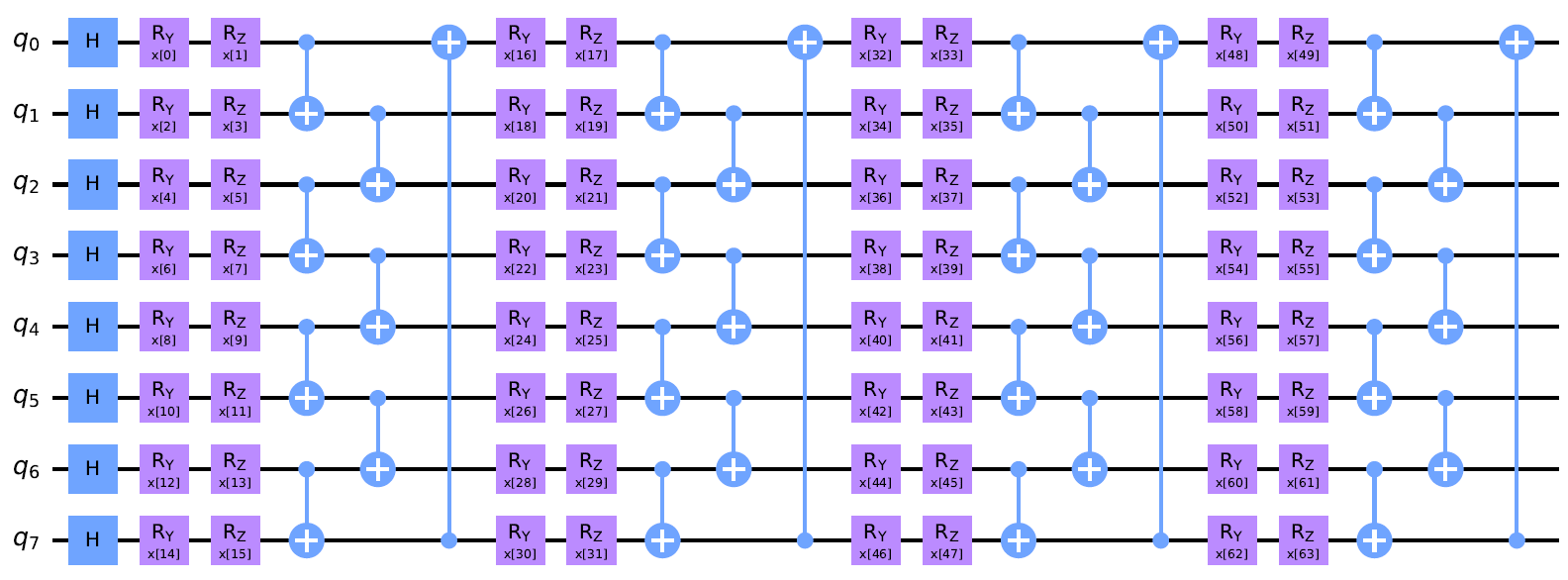}
    \caption{Example of the block encoding method}
    \label{fig:block_enc}
\end{figure}

The circuit will consist of creating blocks of sets of cycling rotation gates (i.e. an x rotation gate
followed by a z rotation gate followed by another x rotation gate). Consecutive rotation gates
must be around different axes. The circuit is created by stacking these blocks together evenly
across all qubits. After a layer of blocks is created a series of CNOT gates are used to connect
consecutive qubit pairs. This process is repeated $L$ times. The rotation amount is defined by the
data itself as was done with the gate encoding above. If there are more gates than data
dimensions the excess rotation gates use 0 for the rotation angle, so they act as pass through
gates. The outputs from this encoding circuit now encode the full data and return unique values
for each of the inputs, without needing to follow a complicated dimensionality reduction
technique. An example of a circuit with eight qubits, four layers, and two gates per block is
shown in \cref{fig:block_enc}.

\subsection{Quantum Machine Learning}

Quantum machine learning is a new and emerging sub-field of quantum computing that
combines two specialized fields into one. The overall process of quantum machine learning is
actually very similar to machine learning on classical computers, since quantum machine
learning is really a hybrid quantum-classical computation \cite{bergholm2018pennylane}. In quantum machine learning a
few key components of the classical machine learning process are replaced by the output of a
quantum computer. Most importantly, the error function to be optimized is at least partially
calculated by a quantum computer. At least one expectation value of a qubit of a quantum circuit
is used to compose the error function \cite{izaac2019basic}, though classical components may be included as well,
which in some cases increases the functionality. For example, in a classification problem the
correct label will be a purely classical value, while the label predicted by the network is
calculated on a quantum computer. Also in order to be able to tune and train the quantum
network, the network must include some classical parameters that can be kept track of and
updated by the algorithm \cite{bergholm2018pennylane,izaac2019basic}.

\begin{figure}[t]
    \centering
    \includegraphics[width=0.95\textwidth]{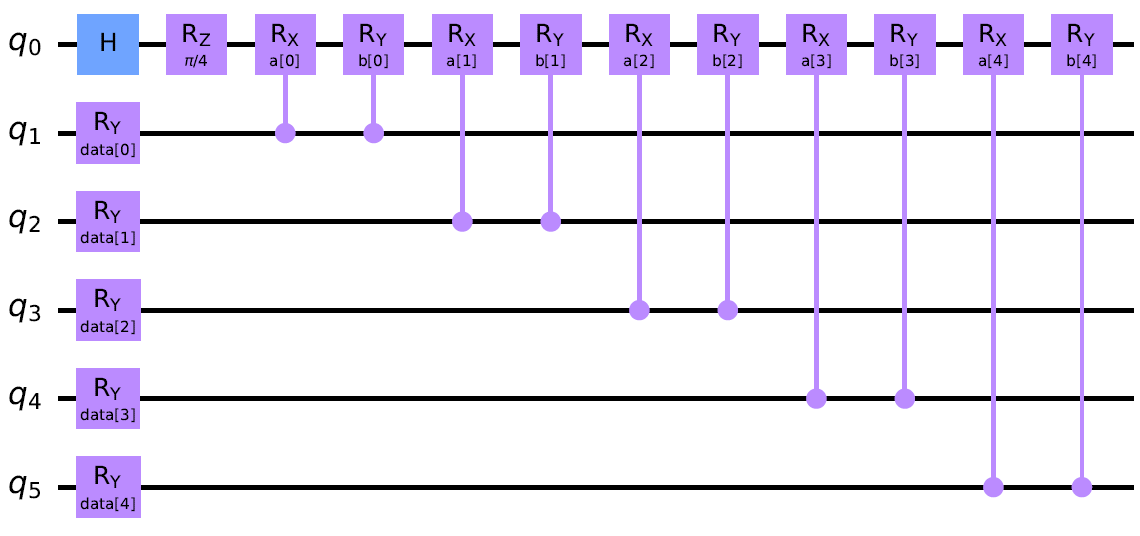}
    \caption{Example quantum machine learning circuit}
    \label{fig:qml_circuit}
\end{figure}

Once the data has been encoded into the quantum circuit using one of the encoding methods
mentioned, multi-qubit operations are applied to the data qubits and the readout qubits with the
goal of manipulating the readout qubits to some desired state corresponding to the data input.
Typically these operations are parametrized controlled rotation gates applied to the readout qubit and controlled off of the data qubits, though non-controlled gates can also be applied to the
readout qubit. Upon running the circuit multiple times, the expectation value of the measured
readout qubit is used as the final output of the circuit and used to compute a loss function. An
example circuit for quantum machine learning is shown in \cref{fig:qml_circuit}. This example used the single
rotation gate data encoding method mentioned above.

In this setup, qubit 0 is the readout qubit and the only one that is measured for the output. The
first two gates acting on qubit 0 place the qubit into an unbiased initial state before the network
operations are applied. Qubits 1-5 are the data qubits. Each rotation operation on each of those
qubits is parameterized by some classical value based on the input data. The remaining gates on
qubit 0 are trainable rotation gates controlled by the state of the data qubits. The network gates
are parameterized by the trainable network variables $a[1], b[1], a[2], b[2], \dots , a[n], b[n]$.

Similar to classical machine learning, the most common method used to find the optimal
parameter values in quantum machine learning is a variant of gradient descent \cite{bergholm2018pennylane}. Since at least
part of the error function is calculated on a quantum computer, the gradient calculation also
requires partial computation on a quantum computer, which leads to another major difference
between classical and quantum machine learning. In classical machine learning fast gradient
calculation is enabled by backpropagation. Backpropagation requires intermediate results to be
measured/calculated and stored for later use to avoid recalculating them many times. Obtaining
intermediate results in the calculation on a quantum computer would require intermediate
measurements. However, on a quantum computer these intermediate measurements would
destroy any quantum behavior being utilized by the quantum computer. This means that in order
to maintain any true quantum calculation, backpropagation is not possible \cite{bergholm2018pennylane,izaac2020quantum} and other methods must be used for calculating the gradient on quantum computers \cite{izaac2020quantum}. Fortunately, other methods of calculating the gradient called parameter-shift rules have been developed and fit very
well into the architecture of quantum computing.


\subsection{Parameter-Shift Rule}

The parameter-shift rule is a very useful tool that allows for an analytically exact gradient
calculation that can be performed on a quantum computer using the same circuit used to calculate the loss, but with shifted parameter values \cite{izaac2020quantum,xanadu2022parameter}. In practice, this results in an
approximate gradient due to the approximation of the expectation value. This is also the case in
the classical machine learning context defined above, however in that case we defined the loss is
the finite approximation (under finite data) to the true expectation value. However, in the
quantum machine learning context, there are in effect two expectations: an expectation with
respect to measurement outcomes and an expectation over the data. In contrast to the finite data
problem, expectation values over measurement outcomes can be run as many times as necessary
to give sufficient precision. 

To show the derivation of this rule, a generic loss function in the form of an expectation value
from a readout qubit will be used. Let the loss function $C(\theta)$ be defined as an expectation value \cite{izaac2020quantum,xanadu2022parameter,schuld2019evaluating}: 
\begin{equation}
    C(\theta) := \bra{\psi} \Uhat_G^\dagger(\theta)\Ahat \Uhat_G(\theta) \ket{\psi}
\end{equation}
where $\ket{\psi}$ is the vector representing the quantum state and $\bra{\psi}$ is its complex conjugate
transpose, $\Uhat_G(\theta)$ is a unitary operator parameterized by $\theta$ with the form $\Uhat_G(\theta)= e^{ia\theta\Ghat}$, where $\Ghat$ is a Pauli operator, $\Uhat_G^\dagger(\theta)$ is the complex conjugate transpose of $\Uhat_G(\theta)$, $\Ahat$ is the observable
being measured, and $a$ is a fixed constant. Taking the derivative with respect to the parameter $\theta$
\begin{equation}
    \derivCtheta = \derivtheta \bra{\psi} \Uhat_G^\dagger(\theta)\Ahat \Uhat_G(\theta) \ket{\psi}
\end{equation}
requires the use of the product rule, giving:
\begin{align}
    \derivCtheta &= \bra{\psi} \derivtheta\big(\Uhat_G^\dagger(\theta)\big)\Ahat \Uhat_G(\theta) \ket{\psi} + \bra{\psi} \Uhat_G^\dagger(\theta)\Ahat \derivtheta \big(\Uhat_G(\theta)\big) \ket{\psi} \\
    &= \bra{\psi} (ia \Ghat) \Uhat_G^\dagger(\theta) \Ahat \Uhat_G(\theta) \ket{\psi} + \bra{\psi} \Uhat_G^\dagger(\theta)\Ahat \Uhat_G(\theta) (-ia \Ghat) \ket{\psi} \\
    &= ia\Big(\bra{\psi} \Ghat \Uhat_G^\dagger(\theta) \Ahat \Uhat_G(\theta) \ket{\psi} - \bra{\psi} \Uhat_G^\dagger(\theta)\Ahat \Uhat_G(\theta) \Ghat \ket{\psi}\Big) \\
    &= ia\bra{\psi} \Uhat_G^\dagger(\theta) \big[\Ghat, \Ahat\big] \Uhat_G(\theta) \ket{\psi} \label{eq:deriv_commutator}
\end{align}
where $\big[\Ghat, \Ahat\big]$ is the commutator $\Ghat \Ahat - \Ahat \Ghat$.  While having a commutator in the calculation seems to complicate things, it does allow for the following identity to be used \cite{xanadu2022parameter,schuld2019evaluating,mitarai2018quantum}:
\begin{equation}\label{eq:comm_identity}
     \big[\Ghat, \Ahat\big] = -i \bigg( \Uhat_G^\dagger\Big(\frac{\pi}{2}\Big)\Ahat \Uhat_G\Big(\frac{\pi}{2}\Big) - \Uhat_G^\dagger\Big(\frac{\pi}{2}\Big)\Ahat \Uhat_G\Big(\frac{\pi}{2}\Big)\bigg),
\end{equation}
where $\Uhat_G(\theta)= e^{ia\theta\Ghat}$, and $\Ghat$ is assumed to be some Pauli operator. Using this identity does limit the application of the final result to be valid only with Pauli operator-based unitary gates; but with how commonly used Pauli gates are, this result is still applicable. For a proof of this identity, see \cref{sec:pauli_comm_proof}.

\begin{figure}[t]
    \centering
    \includegraphics[width=0.9\textwidth]{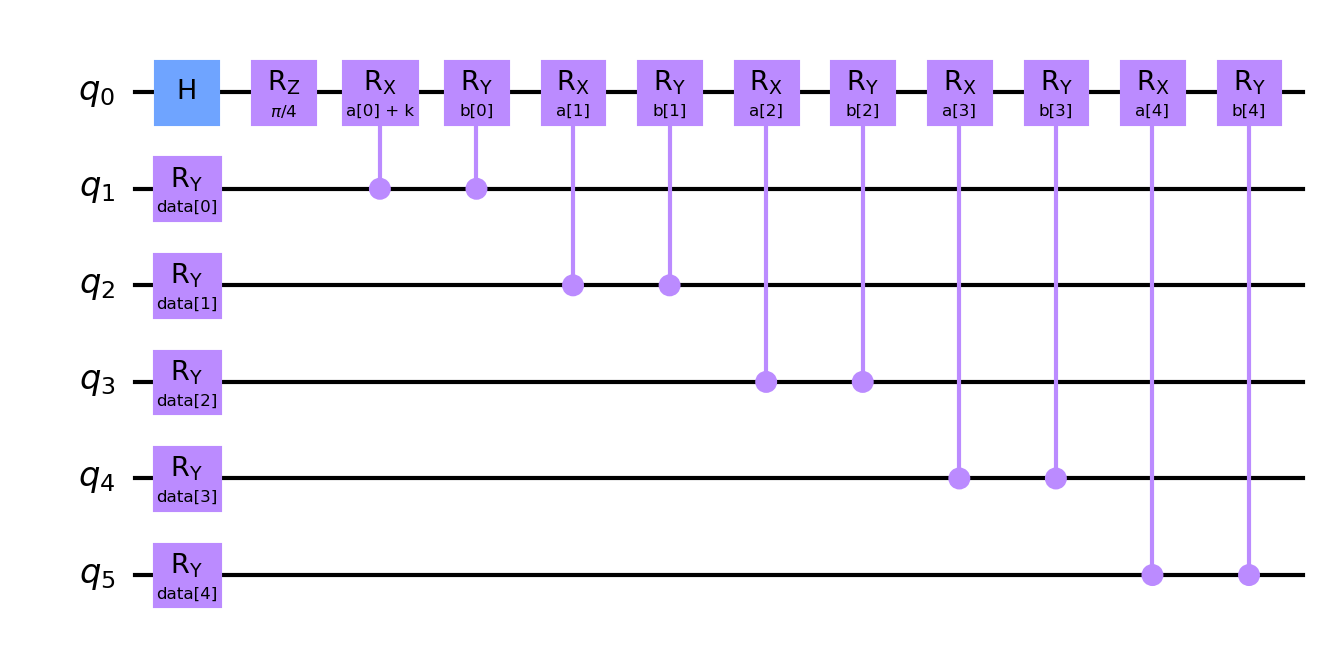}
    \includegraphics[width=0.9\textwidth]{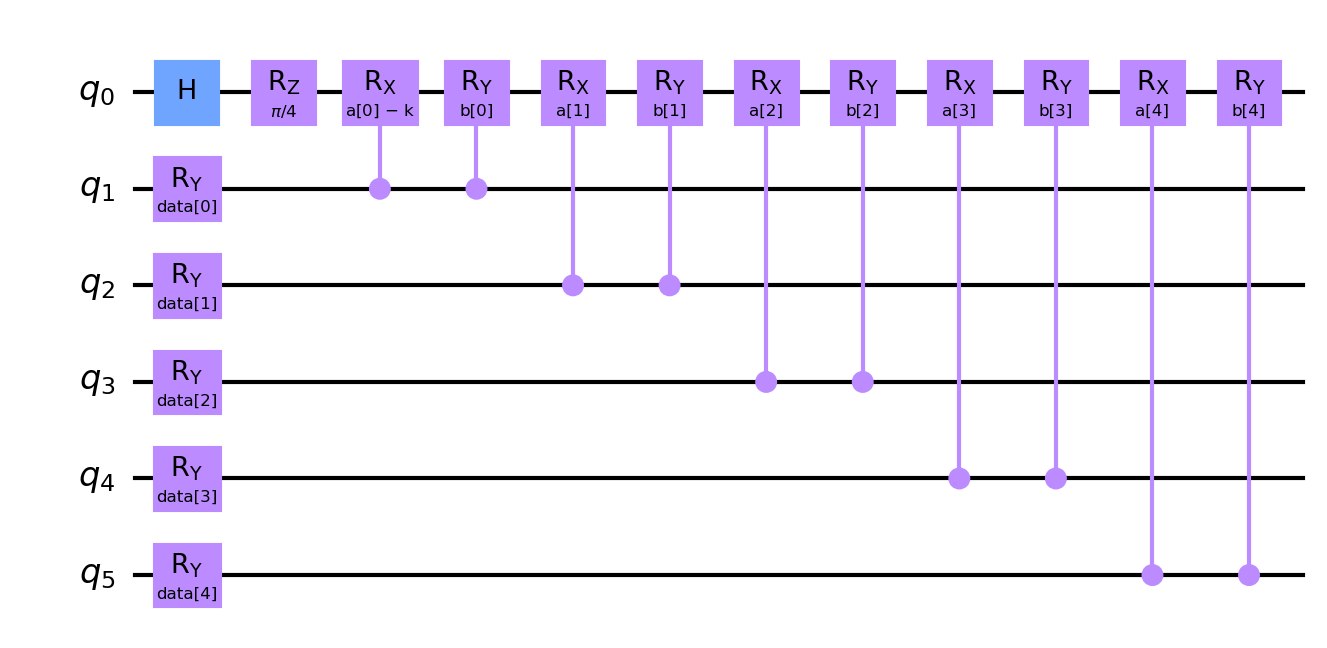}
    \caption{Example circuits for parameter upshift (top) and downshift (bottom) on parameter $a[1]$}
    \label{fig:param_shift}
\end{figure}

Applying this identity to the commutator in \cref{eq:deriv_commutator} leads to a form more
compatible with quantum circuits:
\begin{align}
    \derivCtheta &= ia \bra{\psi}\Uhat_G^\dagger(\theta) (-i) \bigg( \Uhat_G^\dagger\Big(\frac{\pi}{2}\Big)\Ahat \Uhat_G\Big(\frac{\pi}{2}\Big) - \Uhat_G^\dagger\Big(\frac{\pi}{2}\Big)\Ahat \Uhat_G\Big(\frac{\pi}{2}\Big)\bigg)  \Uhat_G(\theta) \ket{\psi} \\
    &= a\bigg( \bra{\psi}\Uhat_G^\dagger(\theta) \Uhat_G^\dagger\Big(\frac{\pi}{2}\Big) \Ahat  \Uhat_G\Big(\frac{\pi}{2}\Big) \Uhat_G(\theta) \ket{\psi} - \bra{\psi} \Uhat_G^\dagger(\theta) \Uhat_G^\dagger\Big(-\frac{\pi}{2}\Big) \Ahat  \Uhat_G\Big(-\frac{\pi}{2}\Big) \Uhat_G(\theta) \ket{\psi}\bigg) \\
    &= a\bigg( \bra{\psi} \Uhat_G^\dagger\Big(\theta + \frac{\pi}{2}\Big) \Ahat  \Uhat_G\Big(\theta + \frac{\pi}{2}\Big) \ket{\psi} - \bra{\psi} \Uhat_G^\dagger\Big(\theta-\frac{\pi}{2}\Big) \Ahat  \Uhat_G\Big(\theta-\frac{\pi}{2}\Big) \ket{\psi}\bigg)  \label{eq:param_shift}
\end{align}
In this form, each term is an expectation value so it can be calculated by a quantum circuit. Of
even more importance to this application is that each term is in the same form as the original loss
function $C(\theta)$ except for the shift by $\pm \pi/2$. This means the gradient calculation can utilize the
exact same circuit as the original loss function. For each parameter’s gradient calculation all that
is required is running the circuit twice more, once with the parameter shifted up by $\pi/2$, and
once with the parameter shifted down by $\pi/2$ \cite{xanadu2022parameter}. Using the circuit given in \cref{fig:qml_circuit} as an
example, to calculate the gradient for the first gate parameter, $a[0]$, and letting $k = \pi/2$, the
circuits in \cref{fig:param_shift} would both be run and the output measured for each circuit.

The difference between the outputs of the two circuits in \cref{fig:param_shift} and the factor of $a$ can be
calculated classically in the hybrid quantum-classical scheme, which will then give the gradient
necessary for gradient descent optimization without requiring intermediate measurements nor
interrupting the quantum behavior of the quantum computer. The gradient calculation process is
repeated for every network parameter, and the parameters are updated according to the gradient
result. With a way to efficiently calculate gradients on a quantum computer, the overall quantum
machine learning process can be described by \cref{alg:param_shift}.

The derivation in this section pertains to gates of the form $\Uhat_G(\theta) = e^{ia\theta \Ghat}$ with $\Ghat$ a Pauli operator. This is fairly general, however there are cases of gates which do not have this operator structure. In such cases, the stochastic parameter-shift rule generalizes the gradient calculation to operators of any form, including multi-qubit gates \cite{killoran2022stochastic}. The interested reader may refer to \cref{sec:stoch_param_shift} for a complete derivation.

While the parameter-shift rule does calculate exact gradients, it suffers from requiring significantly more circuit evaluations than classical gradient descent. Similarly, the stochastic parameter-shift rule also requires multiple circuit evaluations per training iteration. While these methods are very popular methods, they are not the only means of training variational quantum circuits. Other notable algorithms include the Simultaneous Perturbation Stochastic Approximation (SPSA) \cite{spall1998implementation} and the Quantum Natural Gradient (QNG) \cite{stokes2020quantum}.

\begin{algorithm}
\caption{Mini-batch gradient descent algorithm}\label{alg:param_shift}
\begin{algorithmic}
\Ensure Training data $\vX_i,\dots, \vX_m$, training labels $\vy$, learning rate $r$, shift $k$
\State Randomly initialize model parameters $w_1, \dots, w_n$
\For{$i=1,\dots,m$}
    \State Run circuit with parameters $\vw = [w_1,\dots,w_n]$ and calculate the expectation value of the output $\langle \Ahat(\vX_i,\vw)\rangle$
    \State Calculate loss using data label and circuit output $C(\vy_i,\vX_i,\vw) = \vy_i - \langle \Ahat(\vX_i,\vw)\rangle$
    \For{$j=1,\dots,n$}
        \State Upshift $j^{th}$ parameter: $\vw_+ = [w_1,\dots,w_j + k, \dots, w_n]$
        \State Run circuit with new parameter set $\vw_+$ and measure output $\text{upshift} = \langle \Ahat(\vX_i,\vw_+)\rangle$
        \State Downshift $j^{th}$ parameter: $\vw_- = [w_1,\dots,w_j - k, \dots, w_n]$
        \State Run circuit with new parameter set $\vw_-$ and measure output $\text{downshift} = \langle \Ahat(\vX_i,\vw_-)\rangle$
        \State Calculate gradient with respect to $j^{th}$ parameter $\nabla_{w_j} C(\vy_i, \vX_i, \vw) = \frac{1}{a}(\text{upshift} - \text{downshift})$
    \EndFor
    \State Update parameter $w_j = w_j - r\nabla_{w_j} C(\vy_i, \vX_i, \vw)$
\EndFor
\State \Return $\vw$
\end{algorithmic}
\end{algorithm}


\subsection{Variational Quantum Circuit Classification Example -- The XOR Problem}

\begin{figure}[t]
    \centering
    \includegraphics[width=0.5\textwidth]{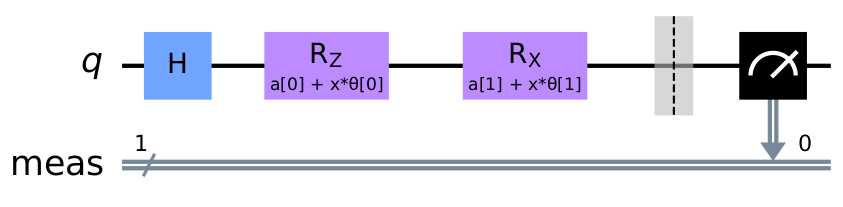}
    \caption{Quantum circuit to solve the XOR problem}
    \label{fig:xor_solution}
\end{figure}

The exclusive-or (XOR) problem was discussed earlier in the introduction to artificial neural
networks. The problem consists of two classes of data on a grid separated into four blocks, where
blocks diagonal from each other contain points in the same class, as depicted in \cref{fig:multilayer_perceptron}. This
results in a classification problem where the two classes are not linearly separable. Comparing
and contrasting the classical and quantum solutions highlights some of the advantages of
quantum computing. Due to the non-linear separation between classes, a classical neural network
requires multiple perceptrons to solve the XOR problem. However, it has been shown that a
simple quantum circuit, shown in \cref{fig:xor_solution}, using only one qubit as a single perceptron can
solve the XOR problem. This approach leverages the phase of the qubit as an extra degree of
freedom \cite{grossu2021single}.

This circuit is fairly simple and consists of only three gates, a Hadamard gate followed by a $Z$-
rotation gate, and then an $X$-rotation gate. The rotation angles of the gates are determined by the
following classical expressions \cite{grossu2021single}:
\begin{align}
    &\text{Z-rotation angle: } \theta_1 x_1 + \alpha \\
    &\text{X-rotation angle: } \theta_2 x_2 + \alpha,
\end{align}
where $\theta_1, \theta_2$ are trainable parameters, $x_1, x_2$ are the input values, and $\alpha$ is another trainable parameter. Here there is a direct solution by using $\theta_1 = \theta_2 = \pi$ and $\alpha = -\pi/2$ \cite{grossu2021single}, however this circuit could be trained by gradient descent. Indeed, for the input vectors $(x_1, x_2) = (0,0)$ and $(x_1, x_2) =(1,1)$ the circuit gives a result near the zero state (up to quantum hard precision), and for the input vectors $(x_1, x_2) = (0,1)$ and $(x_1, x_2) =(1,0)$ the circuit gives an output near the one state (up to quantum hard precision) \cite{grossu2021single}. Finally, with ``noisy" non-integer inputs between 0 and 1, the circuit gives output states between the zero state and the one state (see Table 2 in \cite{grossu2021single}).

\begin{figure}[t]
    \centering
    \includegraphics[width=0.5\textwidth]{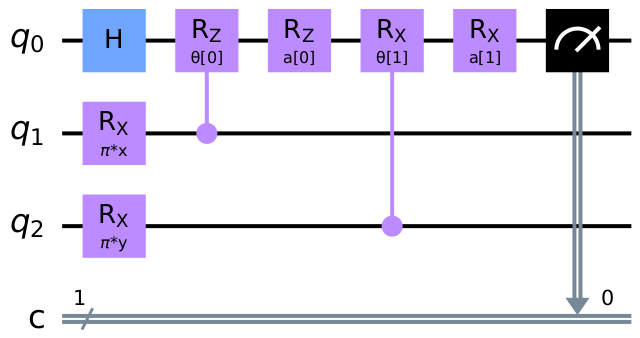}
    \caption{Modified circuit for learning to solve the XOR problem.}
    \label{fig:modified_xor_circuit}
\end{figure}

Extending the circuit and methods introduced above, nearly identical results are obtained using a
quantum machine learning framework and training on a larger data set. To do this, the circuit
from \cite{grossu2021single} is modified to make it more compatible with the parameter-shift gradient descent
method described earlier. This modified circuit is shown below in \cref{fig:modified_xor_circuit}.

This circuit adds two qubits so that the rotation data embedding scheme can be used. The $CRZ(\theta_1)$ gate controlled on qubit 1 for the first data input corresponds to the $\theta_1 x_1$ part of the input parameters to the original $Z$ rotation gate, and similarly for $CRX(\theta_2)$ and $\theta_2 x_2$. Since originally the $\alpha$ parameter is added as a constant, it can be applied in the new circuit as another gate applied in series with the respective controlled gate. Additionally, the single $\alpha$ parameter has been split into $\alpha_1$ and $\alpha_2$ for the Z and X rotations, respectively, to allow the circuit to be more flexible. 

\begin{figure}[t]
    \centering
    \includegraphics[width=0.5\textwidth]{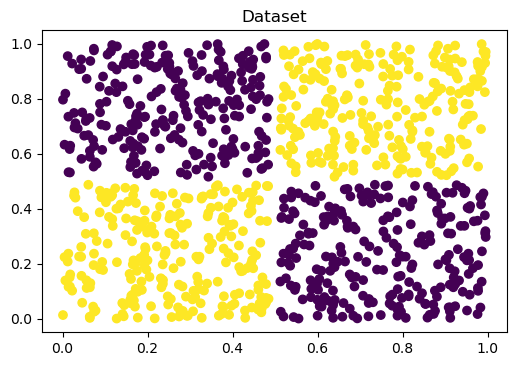}
    \caption{Generated XOR dataset.}
    \label{fig:xor_dataset}
\end{figure}

The data for training and testing this new circuit is generated from a random uniform distribution between 0 and 1 for the $x$ and $y$ values of each data point, though values of exactly 0.5 were excluded as they would be on the class boundary and degenerate. 1000 sample points were generated, with 750 being used for training, 63 used for validation during training, and 187 used for blind testing after training was complete. This dataset is shown in \cref{fig:xor_dataset}.

The yellow data points are assigned to class 0, corresponding to a zero state output of the read-out qubit, and the purple data points are assigned to class 1, corresponding to a one state output of the read-out qubit. The observable used is the Pauli $Z$ gate, which has two possible eigenvalues $\{+1,-1\}$, which are used as the labels for the classes, respectively. The goal for the network is to rotate the read-out qubit towards the zero state for data from class 0, and towards the one state for data from class 1. The expectation value should be closer to +1 for inputs from class 0, and closer to -1 for inputs from class 1. The loss function to be optimized is given by the mean squared error of the expectation value.

To optimize the loss function, mini-batch gradient descent optimization was used, with a batch size of 25 data samples and a learning rate (or step size) of 0.025. The network was trained over 150 epochs. The loss function was averaged over the 25 samples in each mini-batch and that average loss function was used in the gradient calculation. The gradient was calculated using the parameter-shift rule in \cref{eq:param_shift}. To classify a sample in the validation and testing phases of the machine learning process, the continuously valued expectation value output from the network is thresholded. Outputs greater than or equal to  are classified as class 0 and outputs less than  are classified as class 1. 

\begin{figure}[t]%
\centering
\subfigure[]{%
\label{fig:xor_training_loss}%
\includegraphics[width=0.45\textwidth]{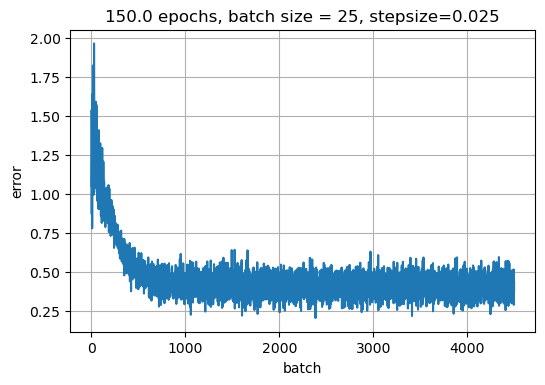}}%
\qquad
\subfigure[]{%
\label{fig:xor_valid_accuracy}%
\includegraphics[width=0.45\textwidth]{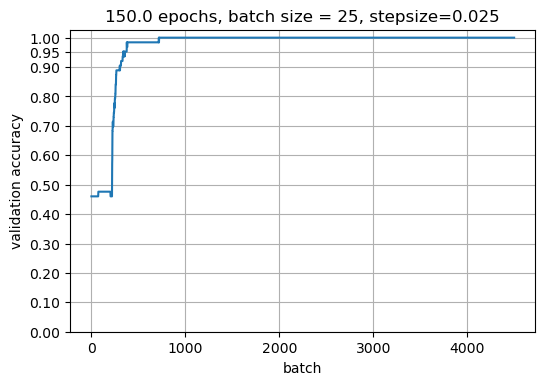}}%
\caption{(a) Training loss vs batch over the training period and (b) Validation accuracy vs batch over the training period.}
\label{fig:xor_training_and_validation}
\end{figure}

\begin{figure}[t]%
\centering
\subfigure[]{%
\label{fig:xor_true_labels}%
\includegraphics[width=0.45\textwidth]{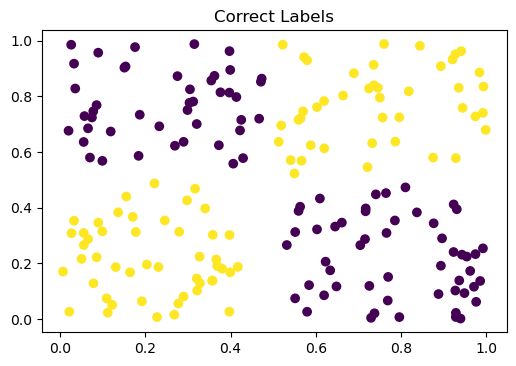}}%
\qquad
\subfigure[]{%
\label{fig:xor_pred_labels}%
\includegraphics[width=0.45\textwidth]{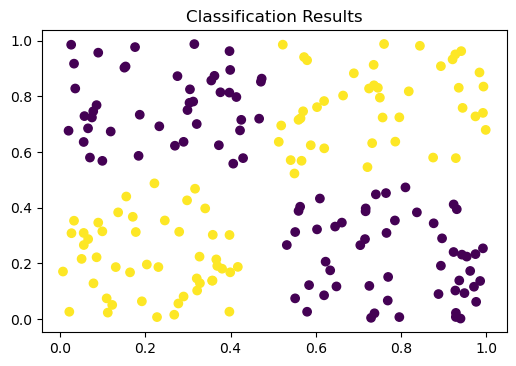}}%
\caption{True (a) and predicted (b) labels for test data.}
\label{fig:xor_true_and_pred_labels}
\end{figure}

\begin{figure}[t]%
\centering
\subfigure[]{%
\label{fig:xor_theta_convergence}%
\includegraphics[width=0.45\textwidth]{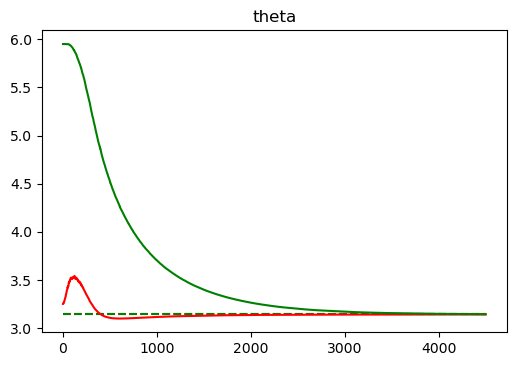}}%
\qquad
\subfigure[]{%
\label{fig:xor_alpha_convergence}%
\includegraphics[width=0.45\textwidth]{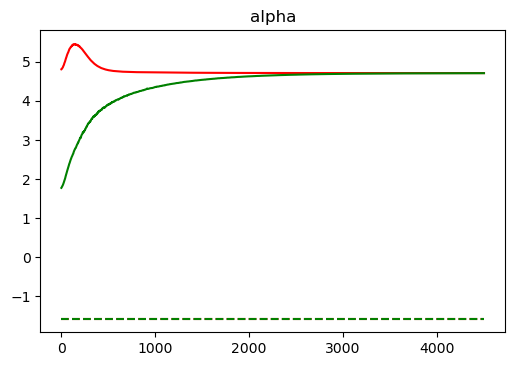}}%
\caption{Parameter convergence for (a) $\theta_1$ and $\theta_2$, and (b) $\alpha$.}
\label{fig:xor_theta_and_alpha_convergence}
\end{figure}

Using these methods, the network was successfully trained and the results found match the results given in \cite{grossu2021single}. The loss was recorded for every batch, and the plot of the loss vs. batch is shown below in \cref{fig:xor_training_loss}. The loss decreased to less than 0.5 on average, and plateaued fairly early in the training process. The plot of validation accuracy per batch over the training period also plateaued early in training as well, as shown in \cref{fig:xor_valid_accuracy}. The validation accuracy converges to 100\% after about 1000 batches, which is an indicator of good network performance. In testing the network performed very well, correctly classifying 100\% of the testing data samples. The true labels and the classification results from the network on the test set are shown in \cref{fig:xor_true_and_pred_labels}. The subplots are identical, showing that the network assigned the true label to every test sample. Furthermore, converged parameters are in full agreement with the parameters used in \cite{grossu2021single}. The convergence of the parameters over the training period are shown below in \cref{fig:xor_theta_convergence} and \cref{fig:xor_alpha_convergence}, where the ideal parameter values from \cite{grossu2021single} are shown by dashed lines.

In \cref{fig:xor_theta_convergence}, the dashed lines are at exactly $\pi$, which correspond to the value used for $\theta_1$ and $\theta_2$ in \cite{grossu2021single}. The network parameters trained here, shown with solid lines, converge to approximately $\pi$. The final values for $\theta_1$ and $\theta_2$ at the end of training were $3.14147$ and $3.14479$, respectively. The results for the $\alpha$ parameters are shown in \cref{fig:xor_alpha_convergence}. The dashed lines are at exactly $-\pi/2$, which correspond to the values used for $\alpha_1$ and $\alpha_2$ in \cite{grossu2021single}. The network parameters $\alpha_1$ and $\alpha_2$ converged to values of $4.71247$ and $4.71022$. 

At first it appears that the network parameters are not in agreement as they converged to different values, however these values are approximately equal to $+3\pi/2$. Since rotations wrap from $2\pi$ back to 0, a rotation by $+3\pi/2$ is equivalent to a rotation by $-\pi/2$. Thus the circuit trained here is equivalent to the circuit presented in \cite{grossu2021single}. These results indicate that quantum circuits can be trained (in simulation) using quantum machine learning methods. Since the XOR problem is an example of a problem that can be solved with a single quantum neuron in contrast to a multi-layer classical perceptron, this simulated demonstration highlights some of the potential advantages of quantum computing and quantum machine learning.

\section{Conclusion}\label{sec:conclusion}

This manuscript introduces the relevant concepts of quantum machine learning, and serves as introductory material. The basic notions of quantum mechanics are described, including quantum phase, superposition, entanglement, and expectations. These are used to introduce quantum gates as fundamental building blocks of the quantum computing framework in comparison with the classical digital logic framework. The basics of classical machine learning are introduced specifically related to deep learning for classification, and are used as a background in order to introduce standard notions in quantum machine learning. Finally these notions are applied to an example problem that highlights some potential advantages of quantum machine learning over its classical counterpart. With the growing capabilities of quantum computers, quantum machine learning holds promise for solving hard problems in a variety of domains, and warrants further investigation into the quantum advantage of quantum machine learning.


\bibliographystyle{ieeetran}
\balance
\bibliography{References}


\newpage
\section*{Appendix}
\beginsupplement

\section{Binary Dimensionality Reduction Example}\label{sec:dim_reduction}

In this example we will assume that our data has a starting dimension of 200 and that we are using a quantum computer/simulator that has 16 qubits. For this example we will assume that we are embedding binary values. To convert the high dimensional floating point data to a 16 bit binary vector, an ensemble of weak classifiers will be used. 

For the weak classifier, the perceptron mentioned in the main document is used. This works well in the ensemble case as the algorithm can be optimized through direct optimization via
\begin{equation}\label{eq:regression_solution}
    \vw = \vy \vX\T(\vX\vX\T)^{-1},
\end{equation}
where $\vy\in \{-1,1\}$ is the vector of true labels and $\vX$ is the set of all training data. This is the solution to solving the equation $\vy = \vw \vX$ for $\vw$.

To train an ensemble of perceptrons the training data is split into $N$ sets of equal size, where $N$ is the number of desired bits in the quantum encoding. A perceptron is then trained, using \cref{eq:regression_solution}, for each of the $N$ sets. The entirety of the data is then passed through each of the $N$ perceptrons. Each $-1$ is converted to zero, and the $N$ outputs are combined together to form the binary representation for the algorithm comparisons.

\section{Statement and Proof of the Pauli Commutator Identity}\label{sec:pauli_comm_proof}

Let $\sigma_i$ be a Pauli matrix, and $B$ any operator. Let $U_i(\theta):= \exp(-i \frac{\theta}{2} \sigma_i)$. Then 
\begin{equation}
[\sigma_i, B] = -i \bigg( U_i^\dagger\Big(\piovertwo\Big) B U_i\Big(\piovertwo\Big) - U_i^\dagger\Big(-\piovertwo\Big)B U_i\Big(-\piovertwo\Big) \bigg)
\end{equation}

\begin{proof}
First, note that for Pauli matrix $\sigma_i$, $i=1,2,3$, $\sigma_i^\dagger = \sigma_i$ (that is, Pauli matrices are Hermitian). Plugging in the definition of $U_i(\piovertwo)$ yields
\begin{equation}
    [\sigma_i, B] = -i \bigg(  \exp\Big(i \pifour \sigma_i\Big) B \exp\Big(-i \pifour \sigma_i\Big) - \exp\Big(-i \pifour \sigma_i\Big)B \exp\Big(i \pifour \sigma_i\Big) \bigg).
\end{equation}
next, we apply the following Pauli matrix identity:
\begin{equation}
    e^{i a (\hat{n}\cdot \vec{\sigma})} = I \cos a + i (\hat{n} \cdot \vec{\sigma}) \sin a, \quad |\hat{n}| = 1,
\end{equation}
where $\vec{\sigma}=\sigma_x \hat{x} + \sigma_y \hat{y} + \sigma_z \hat{z}$. Put simply, this intermediate identity holds through the expansion of the exponential into an infinite sum, splitting the sum into even and odd exponent parts, and noting that $\sigma_i^{2p} = I$, $p\in \mathbb{N}_+$, which causes the cosine (composed of the even exponent terms) to only have prefactor Identity, and the sine (composed of the odd exponent terms) to be left with the Pauli matrix prefactor. In our case, $\sigma_i = \hat{n}_i \cdot \vec{\sigma}$, thus
\begin{equation}\label{eq:pauli_id1}
    \exp\Big(i \pifour \sigma_i\Big) = I \cos \pifour + i \sigma_i \sin \pifour
\end{equation}
Applying this identity, and also the fact that $\cos(-a) = \cos(a)$ and $\sin(-a) = - \sin(a)$, yields
\begin{align}
    [\sigma_i, B] &= -i \bigg(  \Big(I \cos \pifour + i \sigma_i \sin \pifour\Big) B \Big(I \cos \negpifour + i \sigma_i \sin \negpifour\Big) \nonumber \\
    &\qquad\quad - \Big(I \cos \negpifour + i \sigma_i \sin \negpifour\Big)B \Big(I \cos \pifour + i \sigma_i \sin \pifour\Big) \bigg) \\
    &= -i \bigg(  \Big(I \cos \pifour + i \sigma_i \sin \pifour\Big) B \Big(I \cos \pifour - i \sigma_i \sin \pifour\Big) \nonumber \\
    &\qquad\quad - \Big(I \cos \pifour - i \sigma_i \sin \pifour\Big)B \Big(I \cos \pifour + i \sigma_i \sin \pifour\Big) \bigg) \\
    &= -i \bigg(  \Big(I \halfsqrttwo + i \sigma_i \halfsqrttwo\Big) B \Big(I \halfsqrttwo - i \sigma_i \halfsqrttwo\Big) - \Big(I \halfsqrttwo - i \sigma_i \halfsqrttwo\Big)B \Big(I \halfsqrttwo + i \sigma_i \halfsqrttwo\Big) \bigg) \\
    &= -\frac{i}{2} \bigg( B - i B \sigma_i + i \sigma_i B - i^2 \sigma_i B \sigma_i - \Big( B + i B \sigma_i - i \sigma_i B - i^2 \sigma_i B \sigma_i \Big) \bigg) \\
    &= -\frac{i}{2} \bigg( -2 i B \sigma_i + 2i \sigma_i B \bigg) = \sigma_i B - B \sigma_i
\end{align}

\end{proof}

\section{Derivation of the Baker-Campbell-Hausdorff Identity}\label{sec:bch}
Let $\Ahat:\calH \rightarrow \calH$ and $\Bhat:\calH \rightarrow \calH$ be operators in Hilbert space $\calH$, and let $f(\lambda)$ be a single parameter function with $\lambda \in \Cb$, defined by:
\begin{equation}
    f(\lambda) = e^{\lambda \Ahat} \Bhat e^{-\lambda \Ahat}.
\end{equation}
Then
\begin{equation}
    f(\lambda) = e^{\lambda [\Ahat, \cdot]}\Bhat.
\end{equation}

\begin{proof}
Starting with the function $f(\lambda) = e^{\lambda \Ahat} \Bhat e^{-\lambda \Ahat}$, which a typical form of an expectation value in the Heisenberg or Interaction picture, write it as a Taylor series expansion of the form:
\begin{equation}\label{eq:taylor}
    F(x) = \sum_{n=0}^\infty \frac{F^n(a)(x-a)^n}{n!}.
\end{equation}
Taking the first derivative of $f(\lambda)$ yields:
\begin{equation}
    f'(\lambda) = e^{\lambda \Ahat} \Ahat \Bhat e^{-\lambda \Ahat} - e^{\lambda \Ahat} \Bhat \Ahat e^{-\lambda \Ahat} = e^{\lambda \Ahat}[\Ahat, \Bhat] e^{-\lambda \Ahat}.
\end{equation}
The evaluating for $\lambda=0$:
\begin{equation}
    f'(0) = e^{0\Ahat} [\Ahat, \Bhat] e^{-0\Ahat} = [\Ahat, \Bhat] = [\Ahat, \cdot]^1 \Bhat 
\end{equation}
where $[\Ahat, \cdot]^1 \Bhat$ is an alternative notation whose usefulness will become apparent shortly. Repeating the above steps for the second derivative gives:
\begin{align}
    f''(\lambda) &= e^{\lambda \Ahat}\Ahat [\Ahat, \Bhat] e^{-\lambda \Ahat} - e^{\lambda \Ahat}[\Ahat, \Bhat]\Ahat e^{-\lambda \Ahat} = e^{\lambda \Ahat}\big[\Ahat, [\Ahat, \Bhat]\big] e^{-\lambda \Ahat} = e^{\lambda \Ahat} [\Ahat, \cdot]^2 \Bhat e^{-\lambda \Ahat} \\
    f''(0) &= e^{0\Ahat}\big[\Ahat, [\Ahat, \Bhat]\big] e^{-0 \Ahat} = \big[\Ahat, [\Ahat, \Bhat]\big] = [\Ahat, \cdot]^2 \Bhat
\end{align}
Using the developing pattern, the $n^{th}$ derivative evaluated at zero can be written as:
\begin{equation}
    f^n(0) = [\Ahat, \cdot]^n \Bhat.
\end{equation}
Now, writing $f(\lambda)$ as a Taylor expansion \cref{eq:taylor} centered at zero (i.e. with $a=0$) gives:
\begin{equation}
    f(\lambda) = e^{\lambda \Ahat} \Bhat e^{-\lambda \Ahat} = \sum_{n=0}^\infty \frac{f^n(0)(\lambda-0)^n}{n!} = \sum_{n=0}^\infty \frac{[\Ahat,\cdot]^n \Bhat \lambda^n}{n!}.
\end{equation}

The final form of the expression looks like the definition of an exponential,
\begin{equation}
    e^{\lambda \Ahat} = \sum_{n=0}^\infty \frac{ \lambda^n \Ahat^n}{n!}.
\end{equation}

Rewriting $f(\lambda)$ in exponential form yields:
\begin{equation}
    f(\lambda) = e^{\lambda \Ahat} \Bhat e^{-\lambda \Ahat} = \sum_{n=0}^\infty \frac{[\Ahat,\cdot]^n \Bhat \lambda^n}{n!} = \left(\sum_{n=0}^\infty \frac{\lambda^n [\Ahat,\cdot]^n }{n!}\right)\Bhat = e^{\lambda [\Ahat, \cdot]}\Bhat,
\end{equation}
which is the Baker-Campbell-Hausdorff identity.
\end{proof}

\section{Derivation of the Derivative of a Parametric Exponential Operator}\label{sec:exp_deriv}

Let $Z(\theta)$ be a some operator parameterized by $\theta$. Then 
\begin{equation}
\pptheta e^{Z(\theta)} = \int_0^1 e^{(1-s) Z(\theta)}\pptheta Z(\theta) e^{s Z(\theta)}\rd s
\end{equation}

\begin{proof}
First, note that $e^{Z(\theta)} = \sum_{n=0}^\infty \frac{Z(\theta)^n}{n!}$. So 
\begin{align}
    \pptheta e^{Z(\theta)} &= \sum_{n=0}^\infty \frac{1}{n!} \pptheta Z(\theta)^n \\
    &= \pptheta\Big( I + Z(\theta) + \half Z(\theta)^2 +  \dots \Big) \\
    &= 0 +  \pztheta  + \half\Big(\ztheta \pztheta + \pztheta \ztheta \Big) + \nonumber \\
    &\quad + \sixth \Big( \pztheta \ztheta^2 + \ztheta \pztheta \ztheta + \ztheta^2 \pztheta \Big) \nonumber \\
    &\quad + \twefourth \Big( \pztheta \ztheta^3 + \ztheta \pztheta \ztheta^2 + \ztheta^2 \pztheta \ztheta + \ztheta^3 \pztheta \Big) + \dots \\
    &= I \pztheta I + \Big(\half \ztheta + \sixth \ztheta^2 + \twefourth\ztheta^3 + \dots \Big)\pztheta I \nonumber \\
    &\quad + I \pztheta \Big(\half \ztheta + \sixth \ztheta^2  + \twefourth \ztheta^3 + \dots \Big) \nonumber \\
    &\quad + \Big(\twefourth \ztheta^2 + \dots\Big) \pztheta \ztheta + \ztheta \pztheta \Big( \twefourth \ztheta^2 + \dots \Big) + \dots \\
    &= \sum_{n=0}^\infty \sum_{k=0}^n \frac{1}{(n+1)!} \ztheta^k \pztheta \ztheta^{n-k}.
\end{align}
Now, shifting/swapping the indices using the identity \cite{puri2001mathematical}
\begin{equation}
    \sum_{n=0}^\infty \sum_{k=0}^n f_{n,k} = \sum_{k=0}^\infty \sum_{n=k}^\infty f_{n,k} = \sum_{k=0}^\infty \sum_{n=0}^\infty f_{n+k, k},
\end{equation}
we have
\begin{equation}
    \pptheta e^{\ztheta} = \sum_{k=0}^\infty \sum_{n=0}^\infty \frac{1}{(n+k+1)!} \ztheta^k \pztheta \ztheta^n.
\end{equation}
Now multiply by $1=\frac{n!k!}{n!k!}$
\begin{equation}
    \pptheta e^{\ztheta} = \sum_{k=0}^\infty \sum_{n=0}^\infty \frac{n!k!}{n!k!(n+k+1)!} \ztheta^k \pztheta \ztheta^n.
\end{equation}
Next, we note that $\frac{n!k!}{(n+k+1)!}$ is the definition of the beta function, which can be represented as an integral via Euler's beta function identity (proof provided after the conclusion of this proof)
\begin{equation}
    B(k+1, n+1) := \frac{n!k!}{(n+k+1)!} = \int_0^1 (1-s)^k s^n \rd s
\end{equation}
Plugging in Euler's beta function identity yields
\begin{align}
    \pptheta e^{Z(\theta)} &= \int_0^1 \sum_{k=0}^\infty \sum_{n=0}^\infty \frac{1}{n!k!} \Big((1-s)\ztheta\Big)^k \pztheta \Big(s\ztheta\Big)^n \rd s \\
    &= \int_0^1 \sum_{k=0}^\infty \frac{1}{k!} \Big((1-s)\ztheta\Big)^k \pztheta \sum_{n=0}^\infty \frac{1}{n!} \Big(s\ztheta\Big)^n \rd s \\
    &= \int_0^1 e^{(1-s) \ztheta }\pztheta e^{s \ztheta } \rd s
\end{align}
\end{proof}

\section{Statement and Proof of Euler's beta function identity}
\begin{equation}
    B(k, n) := \frac{(n-1)!(k-1)!}{(n+k-1)!} = \int_0^1 (1-s)^{k-1} s^{n-1} \rd s
\end{equation}
\begin{proof}
    This proof is readily available in numerous online sources. First, by integration-by-parts, one can show that for $m \in \Nb$,
\begin{equation}\label{single_int_factorial}
    \int_0^\infty p^{m-1}e^{-p} \rd p = (m-1)!
\end{equation}
Next, consider the product
\begin{align}
    (m-1)!(n-1)! &= \int_0^\infty p^{m-1}e^{-p} \rd p \times \int_0^\infty q^{n-1}e^{-q} \rd q.
\end{align}
Rewriting it as a double integral over the region $x, y \geq 0$ in the $(x,y)$-plane yields
\begin{align}
    (m-1)!(n-1)! = 2\int_0^\infty x^{2m-1}e^{-x^2} \rd x \times 2 \int_0^\infty y^{2n-1}e^{-y^2} \rd y.
\end{align}
Now transform to polar coordinates: $x = r \cos \theta$, $y=r\sin \theta$, and $\rd x \rd y = r\rd r \rd \theta$ 
\begin{align}
    (m-1)!(n-1)! &= 4\int_0^\infty r^{2(m+n-1)}e^{-r^2} r \rd r \times  \int_0^{\pi/2}  \sin^{2n-1} \theta \cos^{2m-1}\theta\rd \theta \\
    &= 2(m+n-1)! \int_0^{\pi/2} \sin^{2n-1} \theta \cos^{2m-1}\theta  \rd \theta,
\end{align}
where we have applied the single integer factorial form in \cref{single_int_factorial}. Thus, we have that
\begin{equation}
    \frac{(m-1)!(n-1)!}{(m+n-1)!} = 2 \int_0^{\pi/2} \sin^{2n-1} \theta \cos^{2m-1}\theta  \rd \theta.
\end{equation}
Finally, make the change of variable $t = sin^2 \theta$, which implies that $\rd t = 2 \sin \theta \cos \theta \rd \theta$ and $\cos^2\theta = 1-t$. Re-writing the integral and applying this change of variables yields
\begin{align}
    2 \int_0^{\pi/2}  \sin^{2n-1} \theta \cos^{2m-1}\theta \rd \theta &=  \int_0^{\pi/2}  \sin^{2n-2} \theta \cos^{2m-2}\theta 2\sin\theta \cos\theta \rd \theta \\
    &= \int_0^1 (1-t)^{n-1} t^{m-1} \rd t
\end{align}
Thus we have concluded that 
\begin{equation}
    \frac{(m-1)!(n-1)!}{(m+n-1)!} = \int_0^1 (1-t)^{n-1} t^{m-1} \rd t
\end{equation}
\end{proof}


\section{Derivation of the Stochastic Parameter-Shift Rule}\label{sec:stoch_param_shift}

The stochastic parameter-shift rule allows for a more generalizable gradient calculation that is
applicable to a wider variety of gates including multi-qubit gates \cite{killoran2022stochastic}. This is done by replacing the operator $\Ghat$ in $\Uhat_G(\theta) = e^{-ia\theta\Ghat}$ with $\Ghat = \Hhat + \theta \Vhat$, where $\Hhat$ is an arbitrary linear combination of Pauli operator tensor products, and $\Vhat$ is a tensor product of Pauli operators. Since multi-qubit operators can be constructed as a sum of tensor products of Pauli operators, the use of $\Hhat$ and $\Vhat$ in this form allows for generalization to arbitrary gates and calculation of the gradient analytically. The loss function $C(\theta)$ then becomes:
\begin{equation}\label{eq:loss_stoch_shift}
    C(\theta) = \bra{\psi} \Uhat_G^\dagger(\theta) \Ahat \Uhat_G(\theta) \ket{\psi} = \bra{\psi} e^{ia\theta(\Hhat + \theta \Vhat)} \Ahat e^{-ia\theta(\Hhat + \theta \Vhat)} \ket{\psi}.
\end{equation}

To find the gradient of this loss function and manipulate it into a form compatible with quantum computers requires several identities. The Baker-Campbell-Hausdorff (BCH) identity \cite{banchi2021measuring} is derived in \cref{sec:bch}, and is given by:
\begin{equation}\label{eq:BCH_identity}
    f(\lambda) = e^{\lambda \Ahat} \Bhat e^{-\lambda \Ahat} = \left(\sum_{n=0}^{\infty}\frac{\lambda^n [\Ahat, \cdot]^n}{n!} \right) \Bhat = e^{\lambda [\Ahat, \cdot]}\Bhat.
\end{equation}

We also apply the commutator identity in \cref{eq:comm_identity}, and the following exponential derivative rule, derived in \cref{sec:exp_deriv} \cite{killoran2022stochastic,banchi2021measuring}:
\begin{equation}\label{eq:exponential_derivative}
    \frac{\partial e^{\Zhat(\theta)}}{\partial \theta} = \int_0^1 e^{(1-s)\Zhat(\theta)} \frac{\partial \Zhat(\theta)}{\partial \theta}e^{s\Zhat(\theta)} \rd s
\end{equation}
With this, we derive an analytic form of the gradient that may still be evaluated by a quantum
computer. The starting point is \cref{eq:loss_stoch_shift}, repeated here for clarity:
\begin{equation*}
    C(\theta) = \bra{\psi} e^{ia\theta(\Hhat + \theta \Vhat)} \Ahat e^{-ia\theta(\Hhat + \theta \Vhat)} \ket{\psi}.
\end{equation*}
Applying the BCH identity \cref{eq:BCH_identity} yields:
\begin{equation}
    C(\theta) = \brapsi e^{ia\HthetaVcomm} \Ahat \ketpsi
\end{equation}
Next, the derivative is taken and passed into the expectation:
\begin{equation}
    \derivCtheta = \partialtheta \brapsi e^{ia\HthetaVcomm} \Ahat \ketpsi = \brapsi \partialtheta e^{ia\HthetaVcomm} \Ahat \ketpsi.
\end{equation}
Next the exponential derivative rule in \cref{eq:exponential_derivative} is applied, where $e^{\Zhat(\theta)} = e^{ia\HthetaVcomm}$ and $\frac{\partial \Zhat(\theta)}{\partial \theta} = \partialtheta ia \HthetaVcomm = ia [\Vhat, \cdot]$, giving:
\begin{equation}
    \derivCtheta = \brapsi \partialtheta e^{ia \HthetaVcomm} \Ahat \ketpsi = \brapsi \int_0^1 e^{(1-s)ia\HthetaVcomm}ia[\Vhat, \cdot] e^{sia\HthetaVcomm}\Ahat \rd s \ketpsi.
\end{equation}
Next, the constants are moved out of the inner product and the BCH identity \cref{eq:BCH_identity} is applied to the term in red:
\begin{align}
    \derivCtheta &= ia \brapsi \int_0^1 e^{(1-s)ia\HthetaVcomm}ia[\Vhat, \cdot] \textcolor{red}{\{e^{sia\HthetaVcomm}\Ahat\}} \rd s \ketpsi \\
    &= ia \brapsi \int_0^1 e^{(1-s)ia\HthetaVcomm}ia[\Vhat, \cdot] e^{sia(\Hhat + \theta \Vhat)}\Ahat e^{-sia(\Hhat + \theta \Vhat)} \rd s \ketpsi \\
    &=  ia \brapsi \int_0^1 e^{(1-s)ia\HthetaVcomm}ia\big[\Vhat,  e^{sia(\Hhat + \theta \Vhat)}\Ahat e^{-sia(\Hhat + \theta \Vhat)} \big] \rd s \ketpsi, 
\end{align}
where we use the commutator notation $[\Ahat, \cdot ] \Bhat = [\Ahat, \Bhat]$. Since the first term of the commutator now only contains $\Vhat$, we can now apply the commutator identity \cref{eq:comm_identity}:
\begin{align}
    \derivCtheta = a \brapsi \int_0^1 e^{(1-s)ia\HthetaVcomm}\bigg(&\Uhat_V^\dagger\Big(\piovertwo\Big)e^{sia(\Hhat + \theta \Vhat)}\Ahat e^{-sia(\Hhat + \theta \Vhat)}\Uhat_V\Big(\piovertwo\Big) \nonumber \\
    & - \Uhat_V^\dagger\Big(-\piovertwo\Big)e^{sia(\Hhat + \theta \Vhat)}\Ahat e^{-sia(\Hhat + \theta \Vhat)}\Uhat_V\Big(-\piovertwo\Big) \bigg) \rd s \ketpsi,
\end{align}
where $\Uhat_V(\theta) = e^{ib\theta \Vhat}$ and $b$ is another constant $b \neq a$. The BCH identity is applied once more to the entire term inside the brackets where
\begin{equation}
    \bigg(\Uhat_V^\dagger\Big(\piovertwo\Big)e^{sia(\Hhat + \theta \Vhat)}\Ahat e^{-sia(\Hhat + \theta \Vhat)}\Uhat_V\Big(\piovertwo\Big) - \Uhat_V^\dagger\Big(-\piovertwo\Big)e^{sia(\Hhat + \theta \Vhat)}\Ahat e^{-sia(\Hhat + \theta \Vhat)}\Uhat_V\Big(-\piovertwo\Big) \bigg)
\end{equation}
is treated as the matrix $\Bhat$ in $e^{\lambda \Ahat} \Bhat e^{-\lambda \Ahat} = e^{\lambda[\Ahat,\cdot]}\Bhat$, and $e^{(1-s)ia\HthetaVcomm}$ is the exponential. This BCH identity application yields:
\begin{align}
    \derivCtheta &= a \brapsi \int_0^1 e^{(1-s)ia(\Hhat + \theta \Vhat)}\bigg(\Uhat_V^\dagger\Big(\piovertwo\Big)e^{sia(\Hhat + \theta \Vhat)}\Ahat e^{-sia(\Hhat + \theta \Vhat)}\Uhat_V\Big(\piovertwo\Big) \nonumber \\
    &\qquad\qquad\qquad\qquad\qquad\quad\; - \Uhat_V^\dagger\Big(-\piovertwo\Big)e^{sia(\Hhat + \theta \Vhat)}\Ahat e^{-sia(\Hhat + \theta \Vhat)}\Uhat_V\Big(-\piovertwo\Big) \bigg) \rd s \ketpsi \\
    &= a \bigg( \brapsi \int_0^1 e^{(1-s)ia(\Hhat + \theta \Vhat)}\Uhat_V^\dagger\Big(\piovertwo\Big)e^{sia(\Hhat + \theta \Vhat)}\Ahat e^{-sia(\Hhat + \theta \Vhat)}\Uhat_V\Big(\piovertwo\Big)  e^{-(1-s)ia(\Hhat + \theta \Vhat)} \rd s \ketpsi \nonumber \\
    &\quad\; - \brapsi \int_0^1 e^{(1-s)ia(\Hhat + \theta \Vhat)}\Uhat_V^\dagger\Big(-\piovertwo\Big)e^{sia(\Hhat + \theta \Vhat)}\Ahat e^{-sia(\Hhat + \theta \Vhat)}\Uhat_V\Big(-\piovertwo\Big)  e^{-(1-s)ia(\Hhat + \theta \Vhat)} \rd s \ketpsi \bigg)\\
    &= a \int_0^1  \bigg( \brapsi e^{(1-s)ia(\Hhat + \theta \Vhat)}\Uhat_V^\dagger\Big(\piovertwo\Big)e^{sia(\Hhat + \theta \Vhat)}\Ahat e^{-sia(\Hhat + \theta \Vhat)}\Uhat_V\Big(\piovertwo\Big)  e^{-(1-s)ia(\Hhat + \theta \Vhat)} \ketpsi \nonumber \\
    &\qquad\quad- \brapsi e^{(1-s)ia(\Hhat + \theta \Vhat)}\Uhat_V^\dagger\Big(-\piovertwo\Big)e^{sia(\Hhat + \theta \Vhat)}\Ahat e^{-sia(\Hhat + \theta \Vhat)}\Uhat_V\Big(-\piovertwo\Big)  e^{-(1-s)ia(\Hhat + \theta \Vhat)} \ketpsi \bigg) \rd s \label{eq:stoch_param_shift}
\end{align}
which is the stochastic parameter-shift rule \cite{killoran2022stochastic,banchi2021measuring}. Using the BCH identity, the gradient was
able to be manipulated back into the form of a difference of two expectation values with the
observable $\Ahat$ at the center, the gate operations and state vector $\ketpsi$ on its right, and their complex conjugates transposed on its left. Comparing \cref{eq:stoch_param_shift} to the original loss function \cref{eq:loss_stoch_shift},
the gradient contains more terms as well as an integral. In order
to perform this gradient calculation, a second quantum circuit that represents the gradient
calculation would need to be set up and run in conjunction with the original circuit used to
calculate the loss \cite{killoran2022stochastic}. Additionally, the integral is approximated by the following sampling
scheme:
\begin{equation}
    \int_0^1 \brapsi F(s) \ketpsi \approx \frac{1}{M}\sum_i^M \brapsi F(s_i) \ketpsi, \quad s_i \sim U(0,1)
\end{equation}
where $U(0,1)$ denotes the uniform distribution on $[0,1]$. A value for $s$ is randomly sampled from a uniform distribution for each run of the circuit \cite{killoran2022stochastic} and expectations are averaged. While this does require more resources to run, it is an accurate quantum calculation of the gradient of the quantum loss function.

\end{document}